%% file: WaveParticle_RevisedNJP.tex
\documentclass[10pt,twocolumn,aps,pra,superscriptaddress,longbibliography,nofootinbib,floatfix,
]{revtex4-2}

\usepackage{times}
\usepackage{graphicx}
\usepackage{dcolumn}
\usepackage{bm}
\usepackage{amsmath}
\usepackage{amssymb}
\usepackage{hyperref}
\hypersetup{colorlinks=true,linkcolor=blue,citecolor=blue,urlcolor=blue}
\usepackage{mathtools}
\usepackage{upgreek}

\begin{document}

\title{Multipath wave-particle duality with exact quantification of which-path knowledge}

\author{L. F. Melo}
\affiliation{Departamento de F\'isica, Universidade Federal de Minas Gerais, Belo Horizonte, MG, Brazil}
\affiliation{QUIIN - Quantum Industrial Innovation, Centro de Compet\^encia Embrapii Cimatec, SENAI CIMATEC, Av.\ Orlando Gomes, 1845, Salvador, BA, 41850-010, Brazil}

\author{O. Jim\'enez}
\affiliation{Centro Multidisciplinario de F\'isica, Vicerrector\'ia de Investigaci\'on, Universidad Mayor, Santiago 8580745, Chile}

\author{L. Neves}
\email{lneves@fisica.ufmg.br}
\affiliation{Departamento de F\'isica, Universidade Federal de Minas Gerais, Belo Horizonte, MG, Brazil}
 
\date{\today}

\begin{abstract}

A quantum system (\emph{quanton}) traverses an interferometer with $N$ equally probable paths and interacts with another quantum system (\emph{detector}) that stores path information in a set of symmetric states. In this interferometric framework, we present a systematic quantitative study of wave-particle duality based on  entropic trade-off relations between quantum coherence, characterized by the relative entropy of coherence of the quanton state, and which-path knowledge, quantified by the mutual information obtained through detector-state discrimination. Unlike prior upper-bound-based formulations, we obtain an exact quantification of which-path knowledge for the discrimination strategies considered here, encompassing previously considered strategies and extending beyond them. A general optimal measurement achieving this is implemented through a state-separation map governed by a continuous parameter $\xi\in[0,1]$, followed by minimum-error (ME) discrimination of either the successful outputs only (standard approach) or both the successful and failure outputs (concatenated approach). Our framework identifies a strategy-dependent hierarchy that generally emerges in the tightness of the resulting duality relations. At $\xi=0$, both approaches reduce to the ME measurement, which typically yields the tightest relation. For $\xi>0$, each approach yields a distinct relation that generally becomes less tight as $\xi$ increases, with the concatenated strategy always yielding a relation at least as tight as its standard counterpart. We further characterize the saturation of these relations by identifying the sets of detector states that lead to it and show that nontrivial saturation occurs only for interferometers with a nonprime number of paths. From these saturating sets, we elucidate how the underlying structure of the quanton-detector correlations gives rise to this behavior and explains the lack of universal saturation in multipath duality relations for pure states.

\end{abstract}

\maketitle

\section{Introduction} \label{sec:Intro}

Wave-particle duality relations are quantitative statements on the trade-off between the ``strength'' of interference effects produced by individual quantum systems (quantons) in an interferometer and the knowledge of which path they took inside it \cite{Wootters1979,Rauch1984,Zeilinger1986,Mittelstaedt1987,Greenberger1988,Mandel1991,Jaeger1995,Englert1996,Schwindt99,Englert2000}. 
In 1996, Englert \cite{Englert1996} derived a relation that embodied the original concept of duality, which relied on actively intervening in the experiment to obtain path information---as conceived in Einstein's recoiling-slit proposal to challenge Bohr's complementarity principle \cite{Bohr1949}. He considered two-path interferometers, inside which the quanton is unitarily coupled to an auxiliary quantum system. As a result, the ancilla stores path information in its states, thereby functioning as a which-path detector. The total amount of information that becomes available in this process is quantified by the distinguishability ($\mathcal{D}$) of the detector states and is extracted by the optimal measurement that discriminates between them. On the other hand, wavelike behavior is quantified by the fringe visibility ($\mathcal{V}$) of the interference pattern obtained after tracing out the detector. In this framework, Englert showed that the trade-off between these quantifiers is 
\begin{align}   \label{eq:DualityEnglert}
\mathcal{V}^2+\mathcal{D}^2 &\leqslant 1,    
\end{align}
where full wavelike behavior ($\mathcal{V}=1$) precludes particlelike behavior ($\mathcal{D}=0$), and vice versa, while also allowing for intermediate cases with partial contributions from both.

Following Englert's formulation for two-path interferometers, subsequent efforts sought to generalize the duality relation to multipath configurations. In this scenario, the increased dimensionality of the quanton Hilbert space and the requirement for discrimination among multiple detector states make the characterization of wavelike and particlelike behaviors significantly more complex. Among the many proposed generalizations \cite{Durr2001,Bimonte2003Comment,Bimonte2003,Bera15,Bagan16,Coles16,Qureshi17,Qureshi19,Bagan20}, a particularly well-established approach was introduced by Bera \textit{et al.}\ \cite{Bera15} and Bagan \textit{et al.}\ \cite{Bagan16}, who derived duality relations between quantum coherence and which-path knowledge. In their approach, the coherence of the quanton state in the path basis characterizes wavelike behavior and is quantified by standard measures such as the $l_1$-norm or the relative entropy of coherence \cite{Baumgratz13}. Conversely, which-path knowledge characterizes particlelike behavior and is quantified by the success probability (or mutual information) associated with detector-state discrimination. 

When the detector states are not orthogonal, perfect discrimination is impossible. Nevertheless, several strategies can optimize the process according to predefined criteria \cite{Chefles00,Barnett09,Bergou10}. If the chosen strategy extracts the total available path information, which-path knowledge coincides with distinguishability; otherwise, it is strictly smaller. Discrimination strategies can often be formulated as an optimization problem in which the error rate is minimized subject to a \emph{fixed rate of inconclusive outcomes} (FRIO) \cite{Chefles98-3,Zhang99,Bagan12}. In this context, Bera \textit{et al.}\ \cite{Bera15} employed an unambiguous discrimination (UD) strategy, characterized by a zero error rate and a nonzero inconclusive rate. Bagan \textit{et al.}\ \cite{Bagan16}, on the other hand, adopted minimum-error (ME) discrimination, characterized by a zero inconclusive rate---the same criterion underlying Englert's derivation of the duality relation in Eq.~(\ref{eq:DualityEnglert}). More recently, Bagan \textit{et al.}\ \cite{Bagan20} generalized their previous analysis by considering a strategy in which the inconclusive rate is minimized for a fixed error rate, a formulation closely related to FRIO \cite{Herzog12}. In all these cases \cite{Bera15,Bagan16,Bagan20}, the authors address a general interferometric framework in which the quanton takes each path with arbitrary probability and the detector states form a generic set of pure nonorthogonal states. As a consequence, the associated detector-state discrimination problem is also general and lacks a closed-form optimal solution, so which-path knowledge has been characterized through upper bounds.

Despite the significant advances achieved within this general framework, some limitations remain. The upper-bound-based characterization of which-path knowledge can overestimate this quantity, preventing a precise assessment of the tightness of the resulting duality relations and their dependence on the employed discrimination strategy. This limitation also obscures the identification of genuine saturation in the relations and hinders a deeper understanding of how quanton-detector correlations shape the trade-off. For instance, upper bounds derived from UD \cite{Bera15} can yield formally saturated relations for pure states, even though the actual which-path knowledge generally fails to reach this bound. 

In this work, we consider a quanton traversing an interferometer with $N$ equally probable paths (a uniform interferometer) and interacting with a detector that stores path information in a set of symmetric states. Within this interferometric framework, we obtain an exact quantification of which-path knowledge for FRIO-based discrimination strategies, encompassing and extending beyond those studied in Refs.~\cite{Bera15,Bagan16,Bagan20}. We thus establish entropic duality relations between quantum coherence and which-path knowledge, quantified respectively by the relative entropy of coherence of the quanton state and the mutual information extracted from the discrimination outcomes. For the resulting family of detector states, the optimal FRIO-based discrimination strategy consists of a state-separation map governed by a continuous parameter $\xi\in[0,1]$ \cite{Prosser16}, followed by ME measurement performed either on the successful outputs only (standard FRIO) or on both the successful and failure outputs (concatenated FRIO). Going beyond the upper-bound-based formulations, our framework identifies a hierarchy that generally emerges in the tightness of the resulting duality relations, determined by the parameter $\xi$ and by the choice between the standard and concatenated approaches. At $\xi=0$, both approaches reduce to the ME measurement, which typically yields the tighter relation. For $\xi>0$, each approach yields a distinct relation that generally becomes less tight as $\xi$ increases, with the concatenated strategy always yielding a relation at least as tight as its standard counterpart. Within this hierarchy, the UD-based relation (standard FRIO at $\xi=1$ for linearly independent states) typically appears as the least tight among the considered strategies, highlighting the limitations of the formulation based on the UD upper bound.

We further characterize the saturation of these relations by identifying the sets of detector states that lead to it. Using the equality conditions for the Fourier entropic uncertainty relation, we show that these states constitute a Fourier basis for an $n$-dimensional subspace of the detector space, where $n$ divides $N$. Consequently, saturation beyond the trivial limits of fully wavelike or fully particlelike behaviors occurs only for interferometers with a nonprime number of paths. From these saturating sets, we also elucidate how the underlying structure of the quanton-detector correlations gives rise to this phenomenon and, in turn, explains the lack of universal saturation for pure states.

The paper is organized as follows. In Sec.~\ref{sec:InterfFramework}, we introduce the interferometric framework and summarize the theoretical tools that underpin our quantitative analysis of wave-particle duality. In Sec.~\ref{sec:entropicduality}, we provide an exact quantification of the which-path knowledge and derive the corresponding duality relations for different discrimination strategies, identifying the hierarchy that generally emerges in their tightness. Section~\ref{sec:NumericalResults} presents numerical simulations that illustrate and compare the resulting relations, highlighting their relative tightness and revealing nontrivial saturation behavior depending on the number of interferometric paths. In Sec.~\ref{sec:saturation}, we characterize the sets of detector states and the corresponding structure of the quanton-detector correlations responsible for the observed nontrivial saturation. Section~\ref{sec:Conclusion} presents our concluding remarks.

\section{Interferometric framework and theoretical tools}
\label{sec:InterfFramework}

In this section, we present the interferometric framework and the theoretical tools for the quantitative analysis of wave-particle duality.

\subsection{Uniform \texorpdfstring{$N$}{N}-path interferometers with symmetric which-path detector states}

Consider a quanton that, upon entering an $N$-path interferometer, is prepared in the superposition state
\begin{align}   \label{eq:psiq}
    |\psi\rangle_q &= \frac{1}{\sqrt{N}}\sum_{\ell=0}^{N-1}|\ell\rangle_q,
\end{align}
where $\{|\ell\rangle_q\}_{\ell=0}^{N-1}$ denotes the orthonormal basis associated with the $N$ possible paths. Since each path is taken with equal probability $1/N$, we refer to this interferometer as \emph{uniform}.

Inside the interferometer, the quanton interacts with an auxiliary quantum system in an $N$-dimensional Hilbert space, which is assumed to be initially in a fixed pure state given by
\begin{align}    \label{eq:alpha0}
    |\alpha_0\rangle_d &= \sum_{k\in\mathcal{I}}a_k|k\rangle_d,
\end{align}
where $\mathcal{I}\subseteq\{0,\ldots,N-1\}$ denotes the support of $|\alpha_0\rangle_d$ in the orthonormal basis $\{|k\rangle_d\}_{k=0}^{N-1}$, with $|\mathcal{I}|=n\leqslant N$; the coefficients $a_k\in \mathbb{R}$ are strictly positive for all $k\in\mathcal{I}$, satisfy $\sum_{k\in\mathcal{I}} a_k^2=1$, and vanish for $k\notin\mathcal{I}$. The interaction is implemented through the controlled unitary operation
\begin{align}
    \hat{U}_{qd} &= \sum_{\ell=0}^{N-1}|\ell\rangle_q\langle\ell|\otimes\hat{V}_d^\ell,
\end{align}
where 
\begin{align}
    \hat{V}_d &= \sum_{k=0}^{N-1}\omega^k|k\rangle_d\langle k|,
\end{align}
with $\omega=\exp(2\pi i/N)$. Thus, the initial bipartite state $|\psi\rangle_q|\alpha_0\rangle_d$ evolves into
\begin{align}  \label{eq:Psi}
    |\Psi\rangle &= \frac{1}{\sqrt{N}}\sum_{\ell=0}^{N-1}|\ell\rangle_q|\alpha_\ell\rangle_d,
\end{align}
where
\begin{align}   \label{eq:symmetric}
    |\alpha_\ell\rangle_d &= \sum_{k\in\mathcal{I}}a_k\omega^{k\ell}|k\rangle_d
\end{align}
are \emph{symmetric} states under the action of $\hat{V}_d$, as they satisfy $|\alpha_\ell\rangle_d=\hat{V}_d|\alpha_{\ell-1}\rangle_d=\hat{V}_d^\ell|\alpha_0\rangle_d$ and $|\alpha_0\rangle_d=\hat{V}_d|\alpha_{N-1}\rangle_d$ \cite{Ban97,Chefles98-2}. Clearly, the correlation established by the interaction encodes information about the quanton's path in the state of the auxiliary system, which thereby acts as a which-path detector. Note that, by construction, all detector states $\{|\alpha_\ell\rangle_d\}_{\ell=0}^{N-1}$ belong to the $n$-dimensional support subspace $\mathrm{span}\{|k\rangle_d\}_{k\in\mathcal I}$ of the full detector Hilbert space.

\subsection{Characterizing wavelike properties}

Recent approaches to wave-particle duality relations in multipath interferometers show that quantum coherence not only quantifies, but also offers a clear and intuitive picture of the quanton's wavelike behavior \cite{Bera15,Bagan16,Qureshi17,Bagan20}. In this context, an entropic quantifier of this property is the relative entropy of coherence, which, for a given state $\hat{\rho}$, is defined as \cite{Baumgratz13}
\begin{align}
    C_r(\hat{\rho}) &= S(\hat{\rho}_\textrm{diag})-S(\hat{\rho}),
\end{align}
where $S(\hat{\varrho})=-\textrm{Tr}(\hat{\varrho}\log \hat{\varrho})$ is the von Neumann entropy\footnote{Throughout this article, all logarithms are base 2 and we adopt the standard convention $0\log 0\equiv 0$, which follows by continuity.} and $\hat{\rho}_\textrm{diag}$ denotes the diagonal part of $\hat{\rho}$ with respect to a reference orthonormal basis. $C_r(\hat{\rho})$ measures the amount of information required to distinguish $\hat{\rho}$ from $\hat{\rho}_\textrm{diag}$, and thus captures contributions associated with the off-diagonal structure of the density matrix, which enables interference effects.

In the path basis, the \textit{a priori} coherence, $C_{r0}$, of the quanton state $|\psi\rangle_q$ in Eq.~(\ref{eq:psiq}) is 
\begin{align}  \label{eq:Cr0q}
    C_{r0}(|\psi\rangle_q\langle\psi|) &= \log N,
\end{align}
which is the maximum value, as expected for a uniform $N$-path interferometer.

After the quanton-detector interaction, the reduced state of each subsystem is obtained from Eq.~(\ref{eq:Psi}) by tracing out the other, i.e., $\hat{\rho}_q = \textrm{Tr}_d|\Psi\rangle\langle\Psi|$ and $\hat{\rho}_d = \textrm{Tr}_q|\Psi\rangle\langle\Psi|$. In particular, the reduced state of the detector is given by
\begin{align}    \label{eq:rho_d}
    \hat{\rho}_d &= 
    \frac{1}{N}\sum_{\ell=0}^{N-1}|\alpha_\ell\rangle_d\langle\alpha_\ell| = \sum_{k\in\mathcal{I}}a^2_k|k\rangle_d\langle k|,
\end{align}
where we used Eq.~(\ref{eq:symmetric}) and the fact that $\omega$ is a primitive $N$th root of unity, which satisfies the orthogonality relation $\sum_{\ell=0}^{N-1}\omega^{\ell(k-k')}=N\delta_{kk'}$. Since $\hat{\rho}_q$ and $\hat{\rho}_d$ are the reduced states of a global pure state, it follows that $S(\hat{\rho}_q)=S(\hat{\rho}_d)$. Therefore, as $[\hat{\rho}_q]_\textrm{diag}=\frac{1}{N}\sum_\ell|\ell\rangle_q\langle\ell|$, the \textit{a posteriori} coherence of the quanton state is
\begin{align}   \label{eq:Crrhoq}
    C_r(\hat{\rho}_q) &= \log N - H(\{a_k^2\}),
\end{align}
where $H(\{a_k^2\})$ denotes the Shannon entropy of the probability distribution $\{a_k^2\}$.

The effective measure we adopt to characterize the quanton's wavelike properties is the normalized version of the coherence in Eq.~(\ref{eq:Crrhoq}), given by
\begin{align}  \label{eq:normalizedC}
    \mathcal{C} &= \frac{ C_r(\hat{\rho}_q)}{\log N}  = 1-\frac{H(\{a_k^2\})}{\log N},
\end{align}
which satisfies $0\leqslant\mathcal{C}\leqslant 1$. For simplicity, we refer to this normalized quantity as coherence. Note that $\mathcal{C}\leqslant 1$, with equality holding if and only if $|\alpha_\ell\rangle_d=|\alpha_0\rangle_d$ for all $\ell$. Thus, the process of encoding path information into the detector states inevitably reduces the \textit{a priori} coherence. In particular, when these states are mutually orthogonal, we have $H(\{a_k^2\}) = \log  N$. In this case, $\mathcal{C}=0$, implying that no interference effects can arise due to the availability of full which-path information in the detector states. 

\subsection{Characterizing particlelike properties}

As described above, the coupling with the quanton stores which-path information in the detector. To gain knowledge about the quanton's path, one must perform a measurement that discriminates the detector states, noting that not all measurements can extract all the available information. In this context, following Refs.~\cite{Bagan16,Bagan20}, we characterize the particlelike properties of the quanton by the mutual information between the detector states and the outcomes of a measurement designed to discriminate them. 

Let $D$ and $M$ be the random variables associated with the detector states and the measurement outcomes, respectively. We can express the mutual information between them as
\begin{align}
    I(M:D) &= H(D)-H(D|M),
\end{align}
where $H(D|M)$ denotes the conditional entropy of $D$ given $M$. The variable $D$ takes values in the set $\{\ell\}_{\ell=0}^{N-1}$, corresponding to the state $|\alpha_\ell\rangle_d$, with probability $1/N$; hence, $H(D)=\log N$. On the other hand, $M$ takes values in the set $\{j\}_{j=0}^{N'-1}$, corresponding to the probabilities $p_j=\textrm{Tr}(\hat{\Pi}_j\hat{\rho}_d)$ of a given $N'$-outcome positive operator-valued measure (POVM) $\mathbf{\Pi}=\{\hat{\Pi}_j\mid\hat{\Pi}_j\geqslant0,\sum_{j=0}^{N'-1}\hat{\Pi}_j=\hat{I}\}$ performed on the detector. Thus, the conditional entropy can be written as 
\begin{align}    \label{eq:HDM}
    H(D|M) & = \sum_{j=0}^{N'-1}p_j H(\{p_{\ell|j}\}),
\end{align}
where 
\begin{align}     \label{eq:p_ellj}
    p_{\ell|j} &\equiv p(\alpha_\ell|j) = \frac{{}_d\langle\alpha_\ell|\hat{\Pi}_j|\alpha_\ell\rangle_d}{N\textrm{Tr}(\hat{\Pi}_j\hat{\rho}_d)}
\end{align}
is the probability that the detector state was $|\alpha_\ell\rangle_d$, given that the outcome $j$ was obtained. The expression for this conditional probability follows from Bayes' rule. 

Similarly to coherence, we define the normalized mutual information as a measure of the quanton's particlelike behavior, given by
\begin{align}  \label{eq:K}
    \mathcal{K}(\mathbf{\Pi}) &= \frac{I(M:D)}{\log N} = 1-\sum_{j=0}^{N'-1}p_j \frac{H(\{p_{\ell|j}\})}{\log N},
\end{align}
which satisfies $0\leqslant\mathcal{K}(\mathbf{\Pi})\leqslant 1$. The notation $\mathcal{K}(\mathbf{\Pi})$ highlights that this quantity---which we refer to as \emph{which-path knowledge} \cite{Schwindt99,Englert2000}---depends on the specific discrimination strategy applied, expressed by the second term on the right-hand side of Eq.~(\ref{eq:K}). 
When the measurement is able to extract all the available information stored in the detector states, $\mathcal{K}(\mathbf{\Pi})$ reaches its maximum value: 
\begin{align}   \label{eq:distinguishability}
    \mathcal{K}_\textrm{max} &= \max_\mathbf{\Pi}\mathcal{K}(\mathbf{\Pi}).
\end{align}
The maximum mutual information over all POVMs is known as the \emph{accessible information}, and in the context of wave-particle duality relations, it quantifies path \emph{distinguishability} \cite{Englert1996,Bagan16}. Except for a few special cases, there is no general analytical solution to the problem of maximizing the mutual information. As a result, when deriving duality relations in multipath interferometers, one is typically restricted to providing bounds for the which-path knowledge \cite{Bera15,Bagan16,Bagan20}. Otherwise, when an analytical solution is unavailable, exact expressions for $\mathcal{K}(\mathbf{\Pi})$ ($\leqslant\mathcal{K}_\textrm{max}$) can still be obtained by adopting physically motivated strategies for discriminating the detector states, which is the approach we follow.

\subsection{Entropic duality relation between quantum coherence and which-path knowledge} 
\label{subsec:entropicduality}

Just as visibility and distinguishability are constrained by a duality relation in two-path interferometers [see Eq.~(\ref{eq:DualityEnglert})], coherence and accessible information satisfy a similar relation, though not in quadratic form \cite{Bagan16,Bagan20}. To see this, we use the Holevo bound \cite{NielsenBook} along with Eqs.~(\ref{eq:rho_d}) and  (\ref{eq:normalizedC}), obtaining
\begin{align}
  \mathcal{K}_\textrm{max} &\leqslant\frac{S(\hat{\rho}_d)}{\log N} = \frac{H(\{a_k^2\})}{\log N}=1-\mathcal{C},
\end{align}
which implies $\mathcal{C}+\mathcal{K}_\textrm{max}\leqslant 1$. Since the quanton state's coherence is fixed by the set of detector states, this entropic wave-particle duality relation is the tightest possible in $N$-path interferometers. When the which-path measurement $\mathbf{\Pi}$ does not attain the accessible path information, i.e., $\mathcal{K}(\mathbf{\Pi})\leqslant \mathcal{K}_\textrm{max}$, the relation becomes less tight and is given by
\begin{align}
    \mathcal{C}+\mathcal{K}(\mathbf{\Pi}) &\leqslant 1.
\end{align}
In Sec.~\ref{sec:entropicduality}, we derive duality relations of this form, based on the measurement strategies introduced next.

\subsection{Strategies for which-path state discrimination}  \label{subsec:QSD}

Within the framework outlined above, the uniformity of the interferometer implies that acquiring which-path knowledge amounts to discriminating among $N$ symmetric, $n$-dimensional pure states ($N\geqslant n\geqslant2$), each prepared with equal \textit{a priori} probability $1/N$. For this family of states, although the solution that yields $\mathcal{K}_\textrm{max}$ remains unknown, optimal measurement strategies are known for several other criteria \cite{Ban97,Chefles98-2,Jimenez11,Bagan12}. We proceed with a brief description of these  strategies.

The optimal FRIO measurement minimizes the error rate in identifying the detector states subject to a fixed rate of inconclusive outcomes \cite{Chefles98-3,Zhang99,Bagan12}. For equiprobable symmetric states, this strategy can be implemented through the following two-step process \cite{Melo23,Melo25}. First, an optimal separation map \cite{Chefles98}, characterized by a prescribed separation parameter $\xi\in[0,1]$, probabilistically transforms the input detector states $\{|\alpha_\ell\rangle_d\}$ into more distinguishable ones with the maximum success probability \cite{Prosser16}
\begin{align}   \label{eq:SuccessRate}
    P_s(\xi) &= \frac{na^2_\textrm{min}}{(1-\xi)na^2_\textrm{min}+\xi},
\end{align}
where $a_\textrm{min}=\min\{a_k\}_{k\in\mathcal{I}}$. In case of failure, the resulting states become less distinguishable. Then, a minimum-error measurement discriminates the successful outputs, yielding conclusive outcomes (whether correct or not), while the failure outputs are discarded, yielding inconclusive outcomes with a fixed rate $P_f(\xi)=1-P_s(\xi)$. This two-step process is represented by an $(N+1)$-outcome POVM 
$\mathbf{\Pi}_{\scriptscriptstyle\mathrm{FRIO}}=\{\hat{\Pi}_0(\xi),\ldots,\hat{\Pi}_{N-1}(\xi),\hat{\Pi}^f(\xi)\}$. The elements $\hat{\Pi}_j(\xi)$ and $\hat{\Pi}^f(\xi)$, associated with the conclusive and inconclusive outcomes, respectively, are given by \cite{Prosser16,Melo25} 
\begin{subequations}  \label{eq:POVM_std}
\begin{align} 
    \hat{\Pi}_j(\xi) &= |\phi_j(\xi)\rangle\langle\phi_j(\xi)|, \\
    \hat{\Pi}^f(\xi) &= \frac{n}{N}\sum_{i,j=0}^{N-1}\langle u_{i}|u_j\rangle|\phi_{i}^f(\xi)\rangle\langle\phi^f_j(\xi)|, 
\end{align}
\end{subequations}
where
\begin{subequations}  \label{eq:u_phi_states}
\begin{align} 
    |u_j\rangle &= \frac{1}{\sqrt{n}}\sum_{k\in\mathcal{I}}\omega^{jk}|k\rangle, \label{eq:uniform} \\
    |\phi_j(\xi)\rangle &= \sqrt{P_s(\xi)}\sum_{k\in\mathcal{I}}g_k(\xi)\omega^{jk}|k\rangle, \\
    |\phi_j^f(\xi)\rangle &= \sqrt{P_f(\xi)}\sum_{k\in\mathcal{I}}h_k\omega^{jk}|k\rangle, 
\end{align}
\end{subequations}
with $g_k^2(\xi)=(1-\xi+\xi/na_k^2)/N$ and $h_k^2=(a_k^2-a^2_\textrm{min})/(1-na^2_\textrm{min})Na^2_k$. One may readily verify that
$\sum_{j=0}^{N-1}\hat{\Pi}_j(\xi)+\hat{\Pi}^f(\xi) = \hat{I}_{\mathcal{I}}$,
where $\hat{I}_{\mathcal{I}}$ denotes the identity operator on the support subspace
$\mathrm{span}\{|k\rangle_d\}_{k\in\mathcal{I}}$. Therefore, the above operators define a POVM on the effective $n$-dimensional Hilbert space supporting the detector states.

The optimal FRIO measurement is a tunable strategy that balances the error and inconclusive rates within desired bounds, encompassing other fundamental strategies as special cases for extreme values of $\xi$. For $\xi=0$, it reduces to the minimum-error (ME) measurement \cite{Ban97}, described by the $N$-outcome POVM $\mathbf{\Pi}_{\scriptscriptstyle\mathrm{ME}}=\{\hat{\Pi}_j\}_{j=0}^{N-1}$, where $\hat{\Pi}_j=\hat{\Pi}_j(0)=\frac{n}{N}|u_j\rangle\langle u_j|$ [see Eqs.~(\ref{eq:u_phi_states})]. For $\xi=1$, it corresponds to the optimal unambiguous discrimination (UD) for linearly independent states ($N=n$) \cite{Chefles98-2}, or to the optimal maximum-confidence (MC) measurement for linearly dependent states ($N>n$) \cite{Jimenez11}. 

In the standard implementation of the optimal FRIO measurement, the failure output states are discarded, leading to inconclusive outcomes. However, when $n>2$, these outputs often carry useful information about the inputs, opening the possibility for further discrimination attempts. Here, we consider the case in which the failure outputs are subsequently discriminated via a ME measurement, referring to this strategy as \emph{concatenated} FRIO.\footnote{This is a slight abuse of language, since the ``concatenated FRIO'' does not produce inconclusive outcomes. However, we adopt this terminology throughout the article to distinguish it from ``standard FRIO'' and also because it effectively conveys the idea that the main strategy (optimal FRIO) is applied first; if it fails, a second strategy (ME) is then used to extract the remaining information.} It is represented by a $2N$-outcome POVM 
$\mathbf{\Pi}_{\mathrm{conc}}=\{\hat{\Pi}_j(\xi),\hat{\Pi}_j^f(\xi)\}_{j=0}^{N-1}$, where $\hat{\Pi}_j(\xi)$ and $\hat{\Pi}_j^f(\xi)$ are the elements associated with a conclusive identification of the input state after successful and failed separations,
respectively, which are given by \cite{Melo25}
\begin{subequations}  \label{eq:POVM_conc}
\begin{align}    
    \hat{\Pi}_j(\xi) &= |\phi_j(\xi)\rangle\langle\phi_j(\xi)|, \\
    \hat{\Pi}_j^f(\xi) &= |\phi^f_j(\xi)\rangle\langle\phi^f_j(\xi)|, 
\end{align}
\end{subequations}
with the states $|\phi_j(\xi)\rangle$ and $|\phi^f_j(\xi)\rangle$ defined in Eqs.~(\ref{eq:u_phi_states}). It is straightforward to verify that $\sum_{j=0}^{N-1}\left[\hat{\Pi}_j(\xi)+\hat{\Pi}_j^f(\xi)\right]
=\hat{I}_{\mathcal I}$.

Thus, the acquisition of which-path knowledge---and its implications for duality relations---are examined as functions of the discrimination strategy, parametrized by $\xi$ and implemented under both standard and concatenated measurement approaches.

\section{Entropic duality relations across discrimination regimes}
\label{sec:entropicduality}

In this section, we derive entropic duality relations in our interferometric framework across the discrimination regimes described above. By exactly determining the which-path knowledge, we analyze the tightness of the resulting families of relations associated with the different measurement approaches and identify the hierarchy that generally emerges among them.

\subsection{Duality relation for standard FRIO measurement and limiting cases} \label{subsec:DualityFRIO}

For the optimal FRIO measurement, the probabilities appearing in Eq.~(\ref{eq:K}) depend on the separation parameter $\xi$ and are computed using Eqs.~(\ref{eq:symmetric}), (\ref{eq:rho_d}), and (\ref{eq:SuccessRate})-(\ref{eq:u_phi_states}). The total probabilities of obtaining a conclusive and an inconclusive outcome are given by $p_j(\xi)=\textrm{Tr}[\hat{\Pi}_j(\xi)\hat{\rho}_d]=P_s(\xi)/N$ and $p_f(\xi)=\textrm{Tr}[\hat{\Pi}^f(\xi)\hat{\rho}_d]=P_f(\xi)$, respectively. From Eq.~(\ref{eq:p_ellj}), the conditional probability distributions corresponding to conclusive and inconclusive outcomes are, respectively,
\begin{subequations}
    \begin{align}
        p_{\ell|j}(\xi) & = \left|\sum_{k\in\mathcal{I}}a_kg_k(\xi)\omega^{k(j-\ell)}\right|^2, \label{eq:p_cond_ell_j}\\
        p_{\ell|f} & = \frac{1}{N},
    \end{align}
\end{subequations}
where the latter indicates that the failure outcome leaves the posterior distribution uniform and therefore carries no which-path information. The distributions $p_{\ell|j}(\xi)$ for different $j$ are related by cyclic shifts in $\ell$, i.e., $p_{\ell|j}(\xi)=p_{\ell-j|0}(\xi)$, where the subtraction $\ell-j$ is modulo $N$. Since the Shannon entropy is invariant under cyclic permutation of the index $\ell$, it follows that $H(\{p_{\ell|j}(\xi)\})=H(\{p_{\ell|0}(\xi)\})$ for all $j=0,\ldots,N-1$. Thus, inserting these results into Eq.~(\ref{eq:K}) yields the which-path knowledge:
\begin{align}   \label{eq:Kfrio}
    \mathcal{K}(\mathbf{\Pi}_{\scriptscriptstyle\mathrm{FRIO}}) &= P_s(\xi) \widetilde{\mathcal{H}}(\{p_{\ell|0}(\xi)\}), 
\end{align}
where we introduced the shorthand notation  
\begin{align}
     \widetilde{\mathcal{H}}(\{x\}) & := 1-\frac{H(\{x\})}{\log N}
\end{align}
for a quantity that will appear frequently, thereby enabling more compact and clearer expressions in what follows.
Using Eq.~(\ref{eq:normalizedC}), the corresponding duality relation is given by
\begin{align}   \label{eq:DualityFRIO}
    \mathcal{C}+\mathcal{K}(\mathbf{\Pi}_{\scriptscriptstyle\mathrm{FRIO}})  &= \widetilde{\mathcal{H}}(\{a_k^2\}) + P_s(\xi)\widetilde{\mathcal{H}}(\{p_{\ell|0}(\xi)\}) \leqslant 1.
\end{align}
For all $\xi\in[0,1]$, this relation is always saturated in the following two cases, which we refer to as trivial saturation points: when the detector states are orthogonal ($a_k^2=1/N$ for all $k$), in which case $\mathcal{K}(\mathbf{\Pi}_{\scriptscriptstyle\mathrm{FRIO}})=1$, or when they are all identical, resulting in $\mathcal{C}=1$. Later we will show that there also exist nontrivial saturation points in specific interferometric scenarios.

The limiting values $\xi=0$ and $\xi=1$ recover, respectively, the ME and UD (or MC) strategies. When the detector states are linearly independent, at $\xi=1$ we have $p_{\ell|0}(1)=\delta_{0\ell}$, and the which-path knowledge reduces to $\mathcal{K}(\mathbf{\Pi}_{\scriptscriptstyle\mathrm{UD}})=P_s(1)=Na^2_\textrm{min}$, in agreement with Ref.~\cite{Bagan20}. At the opposite limit, $\xi=0$, the detector states are discriminated using the ME strategy, yielding
\begin{align} \label{eq:DualityME}
\mathcal{C}+\mathcal{K}(\mathbf{\Pi}_{\scriptscriptstyle\mathrm{ME}}) &= \widetilde{\mathcal{H}}(\{a_k^2\}) + \widetilde{\mathcal{H}}(\{p_{\ell|0}(0)\}) \leqslant 1.
\end{align}
For the equiprobable symmetric states considered here, the corresponding ME POVM is given by $\hat{\Pi}_j=\frac{n}{N}|u_j\rangle\langle u_j|=\hat{\Phi}^{-1/2}|\alpha_j\rangle\langle \alpha_j|\hat{\Phi}^{-1/2}$, where $\hat{\Phi}=\sum_{j=0}^{N-1}|\alpha_j\rangle\langle \alpha_j|$. This measurement, known as the square-root measurement \cite{Hausladen96}, has two important features regarding the maximization of the mutual information. First, it provides a close approximation to the accessible information when the states to be discriminated are equally likely and nearly orthogonal \cite{Hausladen94}. Second, it satisfies the necessary (but not sufficient) condition for attaining the accessible information \cite{Ban97}. Accordingly, the ME discrimination of the detector states yields an extremum---though not necessarily a maximum---of the which-path knowledge, and closely approximates $\mathcal{K}_\textrm{max}$ when ${}_d\langle\alpha_\ell|\alpha_{\ell'}\rangle_d\approx 0$ for all $\ell\neq\ell'$. The equality $\mathcal{K}(\mathbf{\Pi}_{\scriptscriptstyle\mathrm{ME}})=\mathcal{K}_\textrm{max}$ always holds for two-path interferometers with pure detector states, as the optimal measurements for both criteria coincide \cite{Varga24}. 

\begin{figure}[!t]
\centering
\includegraphics[width=.88\columnwidth]{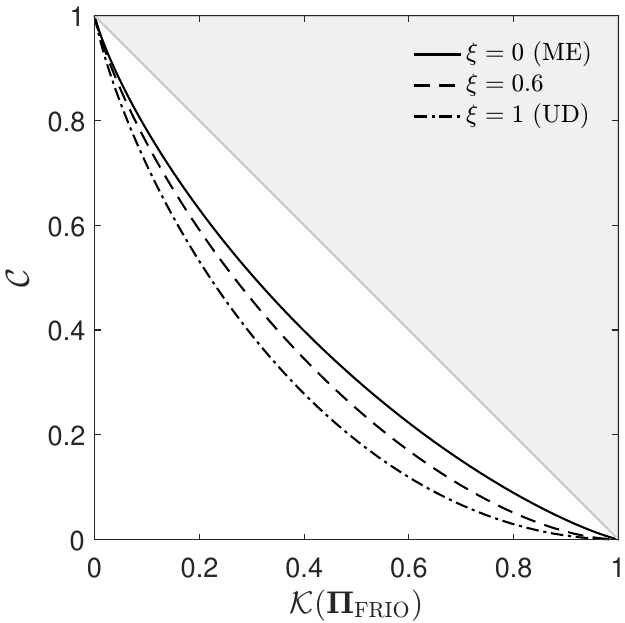}
\caption{Coherence vs which-path knowledge for uniform two-path interferometers with detector states discriminated via optimal FRIO measurement, using the separation parameter $\xi=0$ (solid line, ME measurement), $\xi=0.6$  (dashed line), and  $\xi=1$ (dot-dashed line, optimal UD measurement). The shaded area is the region forbidden by the duality relation.} \label{fig:CxK_N2}
\end{figure}

For $N=2$, it can be demonstrated analytically that the which-path knowledge in Eq.~(\ref{eq:Kfrio}) is a monotonically nonincreasing function of $\xi$, making the duality relation in Eq.~(\ref{eq:DualityFRIO}) progressively less tight as $\xi$ increases. Therefore, the least tight relation occurs at $\xi=1$, which corresponds to the optimal UD (or MC) strategy. This behavior is illustrated in the plot of $\mathcal{C}$ versus $\mathcal{K}(\mathbf{\Pi}_{\scriptscriptstyle\mathrm{FRIO}})$ shown in Fig.~\ref{fig:CxK_N2}, for three values of the separation parameter.

%%%%%%%%%%%%%% FIGURE 1 %%%%%%%%%%%%%%%%%%%%
%%%%%%%%%%%%%%%%%%%%%%%%%%%%%%%%%%%%%%%%

For $N>2$, however, an analogous analytical treatment becomes considerably more involved. We therefore investigated the dependence of $\mathcal{K}(\mathbf{\Pi}_{\scriptscriptstyle\mathrm{FRIO}})$ on $\xi$ numerically for sets of detector states specified by general fiducial coefficients $\{a_k\}$. These tests, discussed in detail in Appendix~\ref{app:MonotonicityFRIO}, show that the monotonic behavior found for $N=2$ does not hold in general. Violations were found for detector states with one dominant fiducial coefficient and $N-1$ approximately degenerate smaller coefficients, with the largest observed violations occurring for the degenerate case, i.e., $\{a_k\}=\{a_\textrm{max},a_\textrm{min},\ldots,a_\textrm{min}\}$, which we refer to as the $1$-max two-amplitude family. For these states, $\mathcal{K}(\mathbf{\Pi}_{\scriptscriptstyle\mathrm{FRIO}})$ can increase over certain intervals of $\xi$, and for the $1$-max two-amplitude family it can even exceed $\mathcal{K}(\mathbf{\Pi}_{\scriptscriptstyle\mathrm{ME}})$. This advantage, however, refers to the comparison between the two strategies for the same detector-state ensemble and does not imply a general superiority of FRIO over ME. 

Thus, although the ordering of the standard FRIO duality relations with $\xi$ is not universal for $N>2$, the overall behavior remains characterized by relations that generally become less tight as $\xi$ increases. Accordingly, ME typically yields the tightest relation, whereas the UD (or MC) limit yields the least tight one. The violations identified above constitute exceptions to this general hierarchy rather than a reversal of the overall trend. Their relevance to the full $\mathcal{C}$--$\mathcal{K}(\mathbf{\Pi})$ regions attainable by the different strategies is examined numerically in Sec.~\ref{sec:NumericalResults}.

\subsection{Duality relation for concatenated FRIO measurement}
\label{subsec:DualityCONC}

Using the POVM defined in Eq.~(\ref{eq:POVM_conc}) and following the same steps as in the standard case, we compute the probabilities appearing in Eq.~(\ref{eq:K}) for the concatenated FRIO measurement.  The total probabilities of obtaining a conclusive outcome after successful and failed events are $p_j(\xi)=P_s(\xi)/N$ and $p_j^f(\xi)=P_f(\xi)/N$, respectively. The corresponding conditional probability distributions $p_{\ell|j}(\xi)$ and $p_{\ell|j}^f$ are given by Eq.~(\ref{eq:p_cond_ell_j}) and 
\begin{align}
    p_{\ell|j}^f & = \left|\sum_{k\in\mathcal{I}}a_kh_k\omega^{k(j-\ell)}\right|^2, 
\end{align}
respectively. Note that the distributions $p_{\ell|j}^f$ do not depend on $\xi$ and, like $p_{\ell|j}(\xi)$, are related by cyclic shifts in $\ell$ for different $j$, i.e., $p_{\ell|j}^f=p_{\ell-j|0}^f$. It follows that $H(\{p_{\ell|j}^f\})=H(\{p_{\ell|0}^f\})$ for all $j=0,\ldots,N-1$. Thus, inserting these results into Eq.~(\ref{eq:K}) and using Eq.~(\ref{eq:Kfrio}), we can write the which-path knowledge as
\begin{align}    \label{eq:Kconc}
    \mathcal{K}(\mathbf{\Pi}_{\mathrm{conc}}) & = \mathcal{K}(\mathbf{\Pi}_{\scriptscriptstyle\mathrm{FRIO}}) +P_f(\xi)\widetilde{\mathcal{H}}(\{p_{\ell|0}^f\}),
\end{align}
which leads to the corresponding duality relation 
\begin{multline}   \label{eq:DualityCONC}
\mathcal{C}+\mathcal{K}(\mathbf{\Pi}_{\mathrm{conc}}) 
= \widetilde{\mathcal{H}}(\{a_k^2\}) 
+ P_s(\xi)\widetilde{\mathcal{H}}(\{p_{\ell|0}(\xi)\}) \\
+ P_f(\xi)\widetilde{\mathcal{H}}(\{p_{\ell|0}^f\}) \leqslant 1.
\end{multline}

Equation~(\ref{eq:Kconc}) immediately implies that $\mathcal{K}(\mathbf{\Pi}_{\mathrm{conc}})\geqslant \mathcal{K}(\mathbf{\Pi}_{\scriptscriptstyle\mathrm{FRIO}})$. Thus, for any fixed detector-state ensemble and separation parameter, concatenation can only increase the which-path knowledge relative to standard FRIO, yielding a tighter duality relation whenever the failure branch carries additional which-path information.

On the other hand, the comparison between $\mathcal{K}(\mathbf{\Pi}_{\mathrm{conc}})$ and $\mathcal{K}(\mathbf{\Pi}_{\scriptscriptstyle\mathrm{ME}})$ is less straightforward. Although ME minimizes the average error probability, thus establishing the natural ordering $P_e^{\scriptscriptstyle\mathrm{ME}}\leqslant P_e^{\mathrm{conc}}$, this does not necessarily imply that $\mathcal{K}(\mathbf{\Pi}_{\scriptscriptstyle\mathrm{ME}})\geqslant \mathcal{K}(\mathbf{\Pi}_{\mathrm{conc}})$. Numerical analysis detailed in Appendix~\ref{app:conc_me} shows that this latter inequality can be violated through two distinct mechanisms: (i) when $\mathcal{K}(\mathbf{\Pi}_{\scriptscriptstyle\mathrm{FRIO}})> \mathcal{K}(\mathbf{\Pi}_{\scriptscriptstyle\mathrm{ME}})$ already holds (inherited violation); and (ii) when $\mathcal{K}(\mathbf{\Pi}_{\scriptscriptstyle\mathrm{FRIO}})\leqslant \mathcal{K}(\mathbf{\Pi}_{\scriptscriptstyle\mathrm{ME}})$, but the additional information recovered from the failure branch is sufficiently large to reverse the inequality (proper violation). In contrast to the standard FRIO case, the proper violations found in our numerical searches were not associated with a single characteristic family of detector states, but occurred for ensembles with different amplitude structures. The observed violations were generally small, with the largest values found being of the order of $10^{-3}$. Their occurrence appears to depend on a nontrivial balance between the different contributions to the which-path knowledge in Eq.~(\ref{eq:Kconc}) rather than on a simple structural property of the detector states.\footnote{A complete characterization of the ensembles producing such violations is beyond the scope of the present work.} 

It is important to distinguish the ensemble-by-ensemble comparison described above from a more global analysis based on the full $\mathcal{C}$--$\mathcal{K}(\mathbf{\Pi})$ regions associated with each measurement strategy. The condition $\mathcal{K}(\mathbf{\Pi}_{\mathrm{conc}})> \mathcal{K}(\mathbf{\Pi}_{\scriptscriptstyle\mathrm{ME}})$ compares the two strategies for the same detector-state ensemble. Nevertheless, as shown numerically in Sec.~\ref{sec:NumericalResults} and further analyzed in Appendix~\ref{app:conc_me}, our results indicate that concatenated FRIO measurements do not extend the upper boundary of the $\mathcal{C}$--$\mathcal{K}(\mathbf{\Pi})$ region attainable with ME within the general symmetric detector-state ensembles considered in this work.

Taken together, these results establish the hierarchy that generally emerges among the duality relations considered here. For a fixed $\xi\in(0,1]$, concatenated FRIO always yields a relation at least as tight as its standard counterpart, while ME ($\xi=0$) generally provides tighter relations than both. The exceptions identified above are restricted to particular detector-state ensembles and do not alter this trend. The resulting overall picture, in which the UD (or MC) limit ($\xi=1$) typically yields the least tight relation, contrasts with the formally saturated relation obtained through the UD upper bound for the which-path knowledge in Ref.~\cite{Bera15}. The present analysis---based on the exact quantification of $\mathcal{K}(\mathbf{\Pi})$---shows that such formal saturation does not, in general, correspond to an attainable limit of the trade-off. In the next section, we further illustrate these findings through numerical simulations.

\section{Comparative analysis of the duality relations via numerical simulations}
\label{sec:NumericalResults}

To assess the wave-particle duality relations derived above and compare their behavior under different measurement strategies, we carry out numerical simulations in \textsc{matlab} according to the following procedure. First, we set the number of paths in the interferometer ($N$) and the dimension of the subspace supporting the detector states ($n$). Then, we generate a given number of random sets of detector states as follows. The amplitudes $a_k$ of the fiducial state [Eq.~(\ref{eq:alpha0})] are obtained from the absolute values of normally distributed random variables and subsequently normalized; the corresponding symmetric detector states are then constructed according to Eq.~(\ref{eq:symmetric}). For each generated set, we compute the quanton state's coherence using Eq.~(\ref{eq:normalizedC}) and the which-path knowledge using either Eq.~(\ref{eq:Kfrio}) or Eq.~(\ref{eq:Kconc}), for a fixed value of the parameter $\xi$. The obtained results are displayed in scatter plots of $\mathcal{C}$ versus $\mathcal{K}(\mathbf{\Pi})$, providing a numerical sampling of the accessible $\mathcal{C}$--$\mathcal{K}(\mathbf{\Pi})$ region. In all subsequent plots, the shaded area corresponds to the region forbidden by the duality relations [Eqs.~(\ref{eq:DualityFRIO}), (\ref{eq:DualityME}), and (\ref{eq:DualityCONC})].

\subsection{Results for the ME measurement}
\label{subsec:ResultsME}

\begin{figure}[!t]
\centering
\includegraphics[width=1\columnwidth]{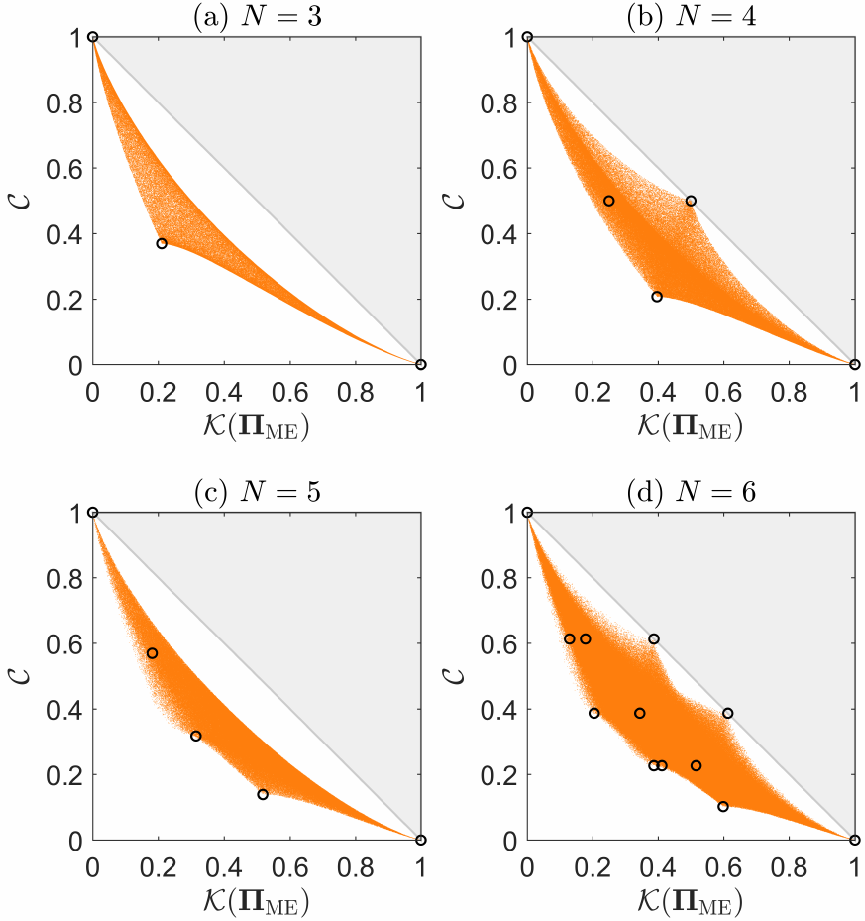}
\caption{Coherence vs which-path knowledge for uniform $N$-path interferometers with detector states discriminated via ME measurement. Orange dots: simulations with (a) $10^5$, (b) $3\times 10^5$, (c) $2.5\times 10^5$, and (d) $3\times 10^6$ randomly generated sets of $N$-dimensional detector states. Open black circles: simulations with all $\binom{N}{n}$ sets of uniform $n$-dimensional detector states; sets yielding the same coherence correspond to a fixed value of $n$, ranging from $n=1$ ($\mathcal{C}=1$) to $n=N$ ($\mathcal{C}=0$). } 
\label{fig:ME}
\end{figure}

Here, we consider detector states discriminated via ME measurement, i.e., $\xi=0$. Following the procedure described above, Fig.~\ref{fig:ME} shows the results for $N=3,4,5,$ and $6$, with $n=N$ (orange dots). In all cases, the scatter plots exhibit the resulting $(\mathcal{K},\mathcal{C})$ points spanning a well-defined region---whose shape varies with $N$---within the bounds imposed by the duality relation. The boundaries of these regions feature cusps and inflection points. We observed empirically that these points correspond to specific sets of \emph{uniform} detector states embedded in $n$-dimensional subspaces of the detector space, i.e., states with $a_k=1/\sqrt{n}$ for all $k\in\mathcal{I}$ [see, e.g., Eq.~(\ref{eq:uniform})]. An indication of this correspondence is that the trivial saturation points $(0,1)$ and $(1,0)$, which are cusps, occur for states of this form with $n=1$ and $n=N$, respectively.

%%%%%%%%%%%%%% FIGURE 2 %%%%%%%%%%%%%%%%%%%%
%%%%%%%%%%%%%%%%%%%%%%%%%%%%%%%%%%%%%%%%

To illustrate the observed correspondence, we first note that for each fixed integer $n\leqslant N$, there exist $\binom{N}{n}$ distinct $n$-dimensional subspaces spanned by subsets of the orthonormal basis $\{|k\rangle\}_{k=0}^{N-1}$. Each subspace, in turn, supports a set of uniform symmetric states. To obtain the corresponding $(\mathcal{K},\mathcal{C})$ points, we generate all $\binom{N}{n}$ such sets for $n$ ranging from $1$ to $N$, with each fixed $n$ yielding $\mathcal{C}=1-\log n/\log N$, and then compute $\mathcal{K}(\mathbf{\Pi}_{\scriptscriptstyle\mathrm{ME}})$. The results are plotted as open black circles in Fig.~\ref{fig:ME}. Among the sets with fixed $n$, many are physically equivalent (inequivalent), yielding the same (different) $\mathcal{K}(\mathbf{\Pi}_{\scriptscriptstyle\mathrm{ME}})$ values and consequently superimposed (displaced) points. As we discuss below, the inequivalent sets are characterized by different structures of the support $\mathcal{I}$ [see Eq.~(\ref{eq:alpha0})], whereas equivalent sets share the same structure. The results in Fig.~\ref{fig:ME} clearly show that the cusps and inflection points of the boundaries arise from sets of uniform detector states.

In particular, for $N=4$ and $N=6$, there are cusps on the saturation line beyond the trivial cases, indicating that the duality relation is also saturated at certain intermediate points. In Sec.~\ref{sec:saturation}, we demonstrate that the relation in Eq.~(\ref{eq:DualityME}) is saturated at nontrivial $(\mathcal{K},\mathcal{C})$ points only for uniform detector states spanning an $n$-dimensional subspace, where $n$ is a nontrivial divisor of $N$ (i.e., $1<n<N$), consistent with the results shown in Fig.~\ref{fig:ME}. In addition, we determine the structure of the support underlying these states.

%%%%%%%%%%%%%% FIGURE 3 %%%%%%%%%%%%%%%%%%%%
\begin{figure}[!t]
\centering
\includegraphics[width=1\columnwidth]{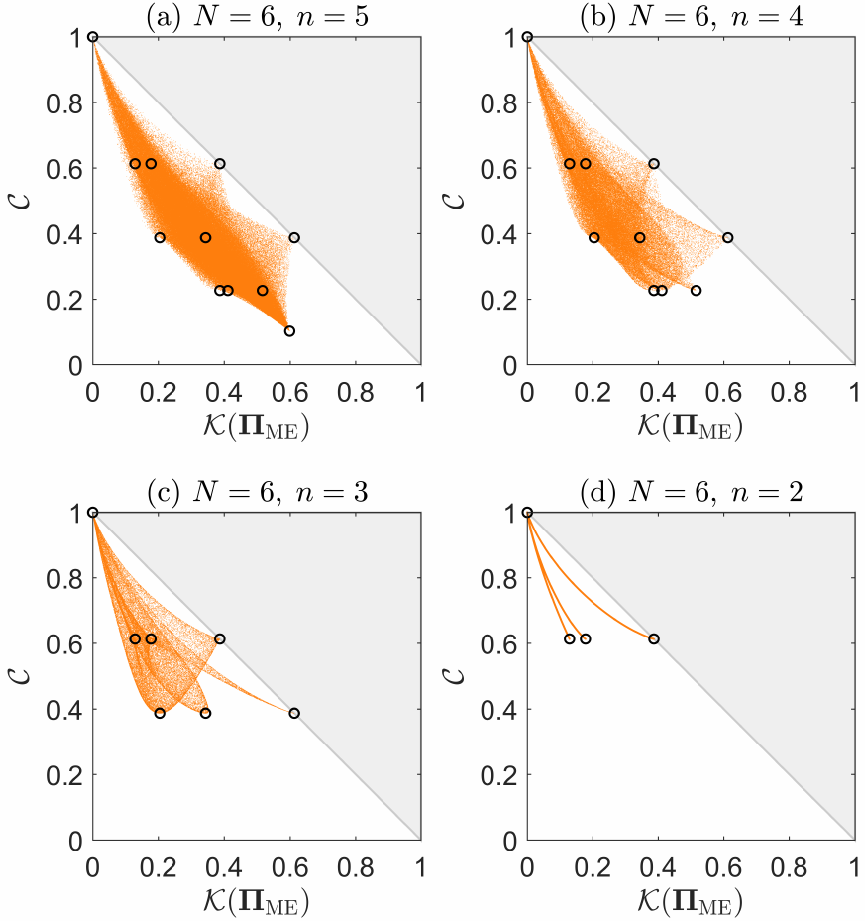}
\caption{Coherence vs which-path knowledge for uniform six-path interferometers with detector states discriminated via ME measurement. Orange dots: simulations with (a) $3\times 10^5$, (b) $10^5$, (c) $5\times 10^4$, and (d) $10^4$ randomly generated sets of $n$-dimensional detector states. Open black circles: simulations with all $\binom{N}{n'}$ sets of uniform $n'$-dimensional detector states; sets yielding the same coherence correspond to a fixed value of $n'$, ranging from $n'=1$ ($\mathcal{C}=1$) to $n'=n$ ($\mathcal{C}=0$). } 
\label{fig:MELD}
\end{figure}
%%%%%%%%%%%%%%%%%%%%%%%%%%%%%%%%%%%%%%%%

Still within the ME framework, in Fig.~\ref{fig:MELD} we present results for a six-path interferometer with linearly dependent detector states spanning $n$-dimensional subspaces, with $n$ ranging from $5$ down to $2$ (orange dots). The only trivial saturation point in this case is $(0,1)$, as the detector states are never orthogonal. Comparing with the plot in Fig.~\hyperref[fig:ME]{\ref{fig:ME}(d)}, we see that the region occupied by the $(\mathcal{K},\mathcal{C})$ points progressively shrinks as $n$ decreases, reflecting the increasingly limited availability of path information. In Fig.~\ref{fig:MELD}, we also plot the $(\mathcal{K},\mathcal{C})$ points for all $\binom{N}{n'}$ sets of uniform $n'$-dimensional detector states, with $n'$ ranging from 1 to $n$ (open black circles). These results show that, in the linearly dependent regime, the uniform $n$-dimensional states yield minima of coherence. However, while coherence is blind to the structure of the support [see Eq.~(\ref{eq:normalizedC})], which-path knowledge is sensitive to it [see Eqs.~(\ref{eq:p_cond_ell_j}) and (\ref{eq:Kfrio})]. Consequently, we observe families of uniform states that, despite yielding the same coherence minima, provide different amounts of path knowledge when extracted via ME measurement. 

To illustrate the role of the support structure on the (in)equivalence of the sets of uniform detector states and how it influences the distribution of the $(\mathcal{K},\mathcal{C})$ points, we examine the case $n=2$ shown in Fig.~\hyperref[fig:MELD]{\ref{fig:MELD}(d)}. In this scenario, three distinct curves emerge from $(0,1)$. The simulations reveal that each curve, from left to right, corresponds to a specific structure of $\mathcal{I}$: unequally spaced nonadjacent ($\mathcal{I}=\{0,2\}$), unequally spaced adjacent ($\mathcal{I}=\{0,1\}$), and equally spaced ($\mathcal{I}=\{0,3\}$), including all their cyclic shifts. A deeper analysis of these aspects is provided in Sec.\ref{sec:saturation}.

\subsection{Results for the standard and concatenated FRIO measurements}

%%%%%%%%%%%%%% FIGURE 4 %%%%%%%%%%%%%%%%%%%%
\begin{figure*}[!t]
    \centering
    \includegraphics[width=.975\textwidth]{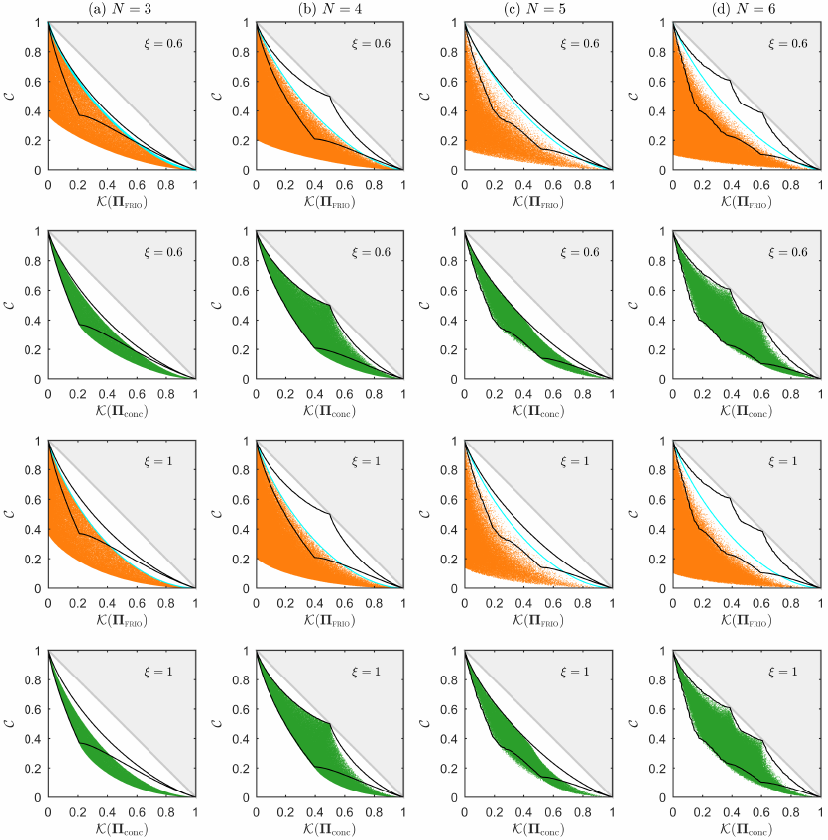}
\caption{Coherence vs which-path knowledge for uniform $N$-path interferometers with $N$-dimensional detector states discriminated via standard FRIO (first and third rows; orange dots) and concatenated FRIO (second and fourth rows; green dots), using the separation parameter $\xi$ shown in the insets. Cyan solid curves for standard FRIO correspond to the $1$-max two-amplitude family of symmetric detector states [Eq.~(\ref{eq:2amp_family})]. The simulations were performed with (a) $10^5$, (b) $3\times 10^5$, (c) $10^5$, and (d) $10^6$ randomly generated sets of detector states. In each plot, the black solid line depicts the numerically estimated boundary of the corresponding $\mathcal{C}$--$\mathcal{K}(\mathbf{\Pi}_{\scriptscriptstyle\mathrm{ME}})$ region shown in Fig.~\ref{fig:ME}.}
\label{fig:CKfrio}
\end{figure*}
%%%%%%%%%%%%%%%%%%%%%%%%%%%%%%%%%%%%%%%%

We now assess the duality relations from Eqs.~(\ref{eq:DualityFRIO}) and (\ref{eq:DualityCONC}), where which-path knowledge is obtained through standard and concatenated FRIO measurements, respectively, using a fixed nonzero value of the separation parameter $\xi$. Again, we consider interferometers with $N=3,4,5$, and $6$ paths, where the quanton is coupled to an $N$-dimensional detector. Figure~\ref{fig:CKfrio} displays plots of $\mathcal{C}$ versus $\mathcal{K}(\mathbf{\Pi}_{\scriptscriptstyle\mathrm{FRIO}})$ (orange dots, first and third rows) and $\mathcal{C}$ versus $\mathcal{K}(\mathbf{\Pi}_{\mathrm{conc}})$ (green dots, second and fourth rows), with columns indicating the value of $N$ and the upper (lower) two rows corresponding to $\xi=0.6$ ($\xi=1$). To facilitate visual comparison with the ME case, we also include numerically estimated boundaries of the corresponding $\mathcal{C}$--$\mathcal{K}(\mathbf{\Pi}_{\scriptscriptstyle\mathrm{ME}})$ regions shown in Fig.~\ref{fig:ME}, depicted here as solid black lines.\footnote{These boundaries were obtained using the $\mathtt{boundary}$ function in \textsc{MATLAB}, which computes a tight polygon enclosing the simulated data points. The shrink factor value used was $0.6$, chosen empirically to provide a faithful visual representation of the outer envelope of the scatter plots in Fig.~\ref{fig:ME}. Note that when the sampling density near the edge of the accessible region is low, the extracted boundary may appear less smooth and slightly underestimate the outer envelope, as observed for $N=5$ and especially $N=6$.}

Taking the ME boundaries as a reference, the $(\mathcal{K},\mathcal{C})$ points for the standard FRIO measurement clearly span a much broader region for all values of $N$ and $\xi$. Overall, as $\xi$ increases, these regions shift away from the saturation line, reflecting the general decrease in tightness discussed in Sec.~\ref{subsec:DualityFRIO}. Since $\mathcal{C}$ is independent of the discrimination strategy applied to the detector states, the observed broadening and overall leftward shift reflect the generally lower which-path knowledge obtained with standard FRIO. The cyan solid curves correspond to the $1$-max two-amplitude family of symmetric states [see Eq.~(\ref{eq:2amp_family})] and trace the upper boundaries of the standard FRIO regions. As discussed in Appendix~\ref{app:MonotonicityFRIO}, this specific family also exhibits the largest observed violations of $\mathcal{K}(\mathbf{\Pi}_{\scriptscriptstyle\mathrm{FRIO}}) \leqslant\mathcal{K}(\mathbf{\Pi}_{\scriptscriptstyle\mathrm{ME}})$. However, these curves lie, to a large extent, within the $\mathcal{C}$--$\mathcal{K}(\mathbf{\Pi}_{\scriptscriptstyle\mathrm{ME}})$ region populated by the general symmetric states, with small excursions observed only at high coherence values, corresponding to nearly parallel detector states.

For the concatenated FRIO measurement, the results in Fig.~\ref{fig:CKfrio} show that the regions spanned by the $(\mathcal{K},\mathcal{C})$ points are more similar in shape and spatially closer to those of the ME case. Thus, compared to standard FRIO, the regions narrow and shift rightward, approaching the saturation line.\footnote{For a fixed $\xi>0$, the only exception to this approach toward the saturation line occurs for $N=3$. In this case, the detector states at the upper boundary of the region belong to the $1$-max two-amplitude family [Eq.~(\ref{eq:2amp_family})], which takes the form $|\alpha_\ell\rangle= a_\textrm{max}|0\rangle+a_\textrm{min}\omega^\ell|1\rangle+a_\textrm{min}\omega^{2\ell}|2\rangle$ and leads to identical failure states in the separation stage ($|\alpha_\ell\rangle \xrightarrow{\textrm{fail}} |0\rangle$ for all $\ell$), making concatenation ineffective.} This behavior follows from the additional information recovered from the failure branch, represented by the second term on the right-hand side of Eq.~(\ref{eq:Kconc}), which ensures that $\mathcal{K}(\mathbf{\Pi}_{\mathrm{conc}})\geqslant \mathcal{K}(\mathbf{\Pi}_{\scriptscriptstyle\mathrm{FRIO}})$. Despite the ensemble-by-ensemble violations of $\mathcal{K}(\mathbf{\Pi}_{\mathrm{conc}})\leqslant \mathcal{K}(\mathbf{\Pi}_{\scriptscriptstyle\mathrm{ME}})$ discussed in Sec.~\ref{subsec:DualityCONC}, the plots provide no visual evidence that concatenated FRIO extends the upper boundary of the $\mathcal{C}$--$\mathcal{K}(\mathbf{\Pi})$ region attainable with ME, consistently with the more detailed numerical analysis presented in Appendix~\ref{app:conc_me}. 

In summary, the results expressed through the $\mathcal{C}$--$\mathcal{K}(\mathbf{\Pi})$ plots of Figs.~\ref{fig:ME}, \ref{fig:MELD}, and \ref{fig:CKfrio} illustrate the behavior of the duality relations derived and discussed in the previous section. They also reveal additional features that only became apparent through graphical analysis. In particular, we highlight the distinctive shapes of the $(\mathcal{K},\mathcal{C})$ point distributions, as well as the emergence of nontrivial saturation points, depending on the specific interferometric scenario. In the next section, we provide a rigorous analysis of the saturation behavior.

\section{On the saturation of the duality relation}
\label{sec:saturation}

In this section, the conditions under which the duality relation is saturated within our interferometric framework are analyzed. We characterize the sets of detector states that lead to saturation (referred to as \textit{saturating sets}) and contrast them with those that do not. Finally, explicit examples are provided to illustrate our findings.

\subsection{Characterizing the saturating sets}

\subsubsection{Fourier entropic uncertainty relation and its equality conditions}

Let $\vec{x}=(x_0,\ldots,x_{N-1})$ be a normalized complex
$N$-dimensional vector, and let $\vec{y}$ denote its
unitary discrete Fourier transform (DFT), with components
$y_\ell=\frac{1}{\sqrt{N}}\sum_{k=0}^{N-1} x_k \omega^{k\ell}$. The corresponding probability distributions $\{|x_k|^2\}$ and
$\{|y_\ell|^2\}$ satisfy the entropic uncertainty
relation \cite{Maassen1988}
\begin{align}
H(\{|x_k|^2\})+H(\{|y_\ell|^2\}) &\geqslant \log_2 N.
\end{align}
Equality in this relation holds if and only if there exist positive integers $n$ and $m$, with $N=nm$, a cyclic shift $\tau$, a phase-modulation index $s$, and a global phase $\phi$ such that \cite{Przebinda2001,Ozaydin2004}
\begin{align}    \label{eq:Equality_condition}
    x_k &=
    \begin{cases}
        \dfrac{e^{i\phi}}{\sqrt{n}}\,\omega^{sk}, &
        k\in\mathcal{I}_{m,\tau}= \{\tau\oplus\kappa m \mid \kappa=0, \ldots, n-1 \}, \\[6pt]
        0, &  \text{otherwise},
    \end{cases}
\end{align}
where $s\in\{0,\ldots,N-1\}$, $\tau\in\{0,\ldots,m-1\}$, and $\tau\oplus\kappa m \equiv (\tau + \kappa m)\!\!\!\pmod N$. Equivalently, up to a cyclic shift, a linear phase modulation, and a global phase, every vector $\vec{x}$ attaining equality has components with uniform magnitude $1/\sqrt{n}$ on an equally spaced support $\mathcal{I}_{m,\tau}$ whose size $n$ divides $N$.

\subsubsection{Saturating sets}

For the duality relation of Eq.~(\ref{eq:DualityME}), where which-path knowledge is obtained via ME measurement, it can be readily shown that saturation occurs if and only if
\begin{align}  \label{eq:Hlambda}
    H(\{a_k^2\})+H(\{|\lambda_\ell|^2\}) &= \log N,
\end{align}
where
\begin{align}  \label{eq:Lambda}
     |\lambda_\ell|^2 & := p_{\ell|0}(0) =  \frac{1}{N}\left|\sum_{k\in \mathcal{I}}a_k\omega^{k\ell}\right|^2,
\end{align}
with $p_{\ell|0}(0)$ obtained from Eq.~(\ref{eq:p_cond_ell_j}). Equation~(\ref{eq:Lambda}) shows that $\{a_k^2\}$ and $\{|\lambda_\ell|^2\}$ are the probability distributions associated with the normalized vector $\vec{a}$ and its unitary DFT $\vec{\lambda}$, respectively. Therefore, the saturation condition in Eq.~(\ref{eq:Hlambda}) is satisfied precisely when the equality in the entropic uncertainty relation for these distributions is reached.

To determine the saturating sets, we apply the equality conditions expressed by Eq.~(\ref{eq:Equality_condition}) to the vector $\vec{a}$, whose components are real and satisfy $a_k>0$ for $k\in\mathcal{I}$ and $a_k=0$ for $k\notin\mathcal{I}$. These conditions require $N=nm$ and  $\mathcal{I}=\mathcal{I}_{m,\tau}$. Moreover, any admissible phase modulation must be constant over the support and can therefore be absorbed into the global phase.\footnote{From Eq.~(\ref{eq:Equality_condition}), the nonzero components of $\vec{a}$ are given by $a_{\tau\oplus\kappa m} =e^{i\phi}\omega^{s(\tau\oplus\kappa m)}/\sqrt{n}$. For successive values of $\kappa$, their ratio is $e^{2\pi i s/n}$. Since these coefficients are real and strictly positive, this ratio must equal unity. Consequently, $\omega^{s(\tau\oplus\kappa m)}=\omega^{s\tau}$ for every $\kappa=0,\ldots,n-1$, and the phase modulation can be absorbed into the global phase. } Consequently, the components of $\vec{a}$ take the form
\begin{align}   \label{eq:aksolution}
    a_k &=
    \begin{cases}
        \dfrac{1}{\sqrt{n}},
        & k\in\mathcal{I}_{m,\tau},\\[6pt]
        0,
        & \text{otherwise}.
    \end{cases}
\end{align}

The unitary DFT of this vector is given by
$\lambda_\ell = \frac{\omega^{\tau \ell}}{\sqrt{Nn}} \sum_{\kappa = 0}^{n-1} \omega^{\kappa m \ell}$, where the sum is a geometric series with ratio $r=\omega^{m\ell}$. Since $r^n=e^{2\pi i\ell}=1$ for all $\ell\in\{0,\ldots,N-1\}$, and $r=1$ if and only if $\ell$ is a multiple of $n$, the sum equals $n$ if $\ell\in\{0,n,2n, \ldots,(m-1)n\}$, and is zero otherwise. Substituting this into $\lambda_\ell$ and taking its modulus squared, we obtain \begin{align}   \label{eq:lambdasolution}
    |\lambda_\ell|^2 &=
    \begin{cases}
        \dfrac{n}{N}, & \ell \in \{0,n,2n, \ldots,(m-1)n\}, 
        \\[6pt]
        0, & \text{otherwise.} 
    \end{cases}
\end{align}
In this case, the entropies of the corresponding probability distributions take the values $H(\{a_k^2\})=\log n$ and $H(\{|\lambda_\ell|^2\})=\log N/n$, whose sum satisfies the saturation condition of Eq.~(\ref{eq:Hlambda}). For any vector $\vec{a}$ distinct from that of Eq.~(\ref{eq:aksolution})---either uniform but supported on a set $\mathcal{I}\neq\mathcal{I}_{m,\tau}$ or nonuniform---the saturation condition cannot be fulfilled.

\begin{figure}[!t]
\centering
\includegraphics[width=1\columnwidth]{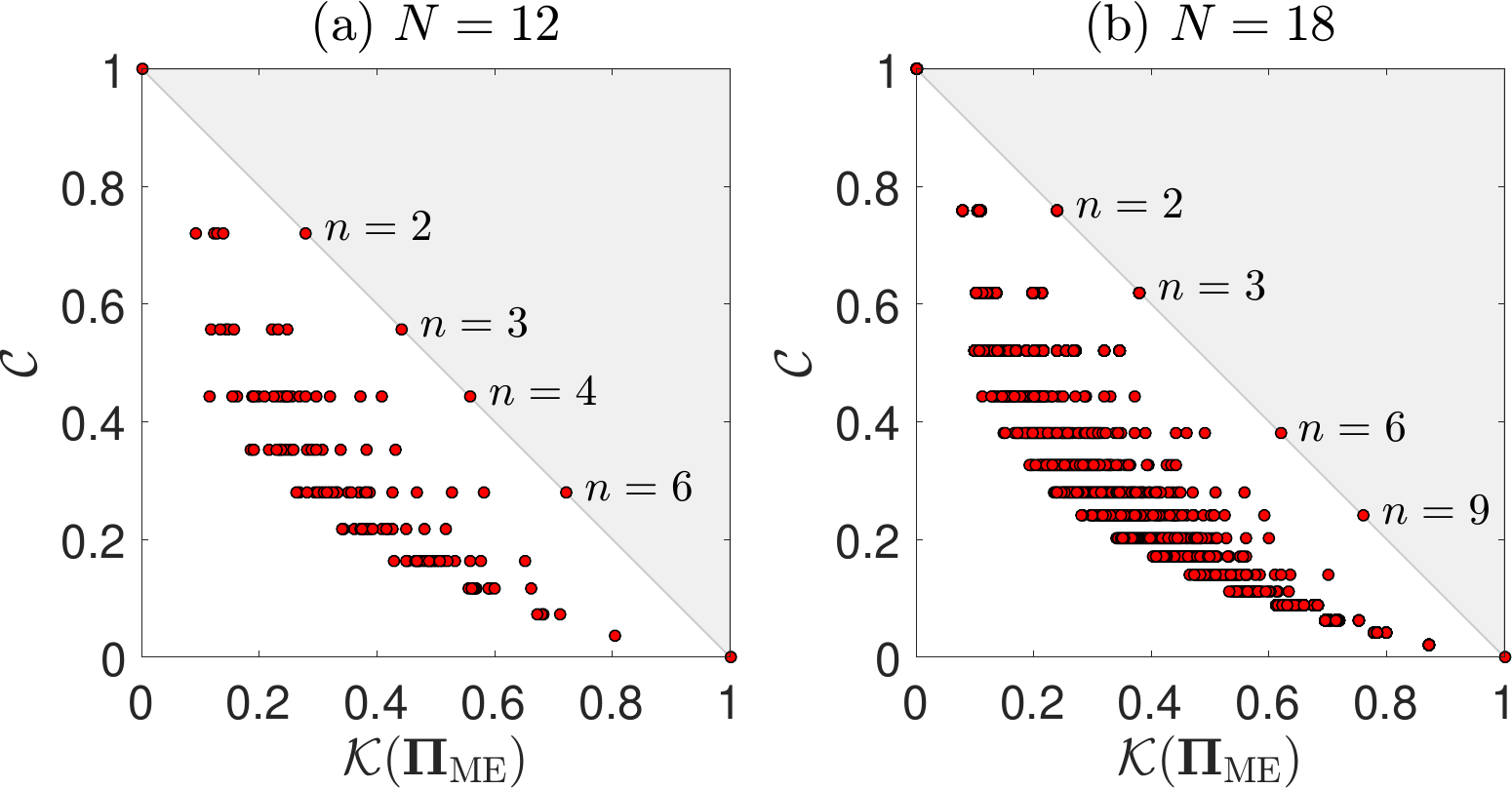}
\caption{Coherence vs which-path knowledge for uniform $N$-path interferometers with uniform $n$-dimensional detector states discriminated via ME measurement. All $\binom{N}{n}$ sets of uniform detector states are generated, and those yielding the same coherence correspond to a fixed value of $n$, ranging from $n=1$ ($\mathcal{C}=1$) to $n=N$ ($\mathcal{C}=0$). The insets show the values of $n$ (beyond the trivial $n=1$ and $n=N$) for which the duality relation of Eq.~(\ref{eq:DualityME}) is saturated.} \label{fig:MEuniform}
\end{figure}

Thus, from Eq.~(\ref{eq:aksolution}), the detector states forming the saturating sets in our interferometric framework are given by
\begin{align}  \label{eq:uniform_mtau}
    |u_\ell^{m,\tau}\rangle_d &= \frac{1}{\sqrt{n}}\sum_{k\in\mathcal{I}_{m,\tau}}\omega^{k\ell}|k\rangle_d  \nonumber \\
    &= \frac{1}{\sqrt{n}}\sum_{\kappa=0}^{n-1}\nu^{\kappa \ell}|\tau\oplus \kappa m\rangle_d,
\end{align}
where $\nu=\exp(2\pi i/n)$; note that an irrelevant global phase factor, $\omega^{\tau\ell}$, has been omitted. As these states are uniform, both the standard and concatenated FRIO strategies reduce to the ME measurement, since no further separation is possible. More importantly, as they span a subspace of dimension $n=N/m$, saturation can occur only when $n$ divides $N$. Consequently, nontrivial saturation points exist exclusively in interferometers with a nonprime number of paths, located at $\eta(N)-2$ positions, where $\eta(N)$ denotes the number of positive divisors of $N$, as illustrated in the examples of Fig.~\ref{fig:ME}. To further exemplify this result, in Fig.~\ref{fig:MEuniform} we show plots of $\mathcal{C}$ versus $\mathcal{K}(\mathbf{\Pi}_{\scriptscriptstyle\mathrm{ME}})$ for interferometers with $N=12$ and $18$ paths, considering only uniform detector states. As already done in Sec.~\ref{subsec:ResultsME}, we compute the values of coherence and which-path knowledge for all $\binom{N}{n}$ sets of uniform states, with $n$ ranging from $1$ to $N$. The obtained results confirm the above description.

%%%%%%%%%%%%%% FIGURE 5 %%%%%%%%%%%%%%%%%%%%
%%%%%%%%%%%%%%%%%%%%%%%%%%%%%%%%%%%%%%%%

The states $|u_\ell^{m,\tau}\rangle_d$ in the saturating sets are characterized by an equally spaced support $\mathcal{I}_{m,\tau}$ in the orthonormal basis $\{|k\rangle_d\}_{k=0}^{N-1}$, with the parameter $m$ setting the spacing and $\tau$ specifying the cyclic shift [see Eq.~(\ref{eq:aksolution})]. Thus, for each $n=N/m$, there exist $m$ physically equivalent saturating sets. In addition, each of these sets forms a Fourier basis for the $n$-dimensional subspace spanned by the vectors $\{|\tau\oplus km\rangle_d\}_{k=0}^{n-1}$. To see this, note that the phase factor $\nu$ in Eq.~(\ref{eq:uniform_mtau}) represents a primitive $n$th root of unity, implying that $\nu^{jn}=1$ and $\sum_{k=0}^{n-1}\nu^{k(\ell-\ell')}=n\delta_{\ell\ell'}$.  As a consequence, 
\begin{align}    \label{eq:periodicity}
    |u_{\ell\oplus n}^{m,\tau}\rangle_d &= |u_{\ell}^{m,\tau}\rangle_d \quad\text{and}\quad {}_d\langle u_{\ell'}^{m,\tau}|u_{\ell}^{m,\tau}\rangle_d=\delta_{\ell\ell'},
\end{align}
where $\ell\oplus n\equiv(\ell+n)\!\!\!\pmod n$. Thus, the states in the saturating sets repeat every $n$ steps in the index $\ell$ and form the Fourier basis $\{|u_{\ell}^{m,\tau}\rangle_d\}_{\ell=0}^{n-1}$ in the corresponding $n$-dimensional subspace.

\subsection{Structure of the quanton-detector correlations}

Using the periodicity of $|u_{\ell}^{m,\tau}\rangle_d$ [see Eq.~(\ref{eq:periodicity})] and reindexing $\ell$ as $\ell=l+jn$, where $l=0,\ldots,n-1$ and $j=0,\ldots,m-1$, the quanton-detector state [Eq.~(\ref{eq:Psi})] for the saturating sets can be expressed as
\begin{align}    \label{eq:Psi_saturation}
    |\Psi^{m,\tau}\rangle &= \frac{1}{\sqrt{n}}\sum_{l=0}^{n-1}|\psi_l\rangle_q|u_{l}^{m,\tau}\rangle_d, 
\end{align}
where
\begin{align}  \label{eq:psi_lq}
    |\psi_l\rangle_q &= \frac{1}{\sqrt{m}}\sum_{j=0}^{m-1}|l+jn\rangle_q, 
\end{align}
with ${}_q\langle\psi_{l'}|\psi_l\rangle_q=\delta_{ll'}$. Thus, $|\Psi^{m,\tau}\rangle$ is a maximally entangled state of Schmidt rank $n$, exhibiting a block structure in the path basis: each quanton component $|\psi_l\rangle_q$ corresponds to a uniform superposition of $m$ path states whose indices differ by multiples of $n$, and is correlated with the detector state $|u_l^{m,\tau}\rangle_d$, which belongs to an orthonormal set. Consequently, the ME measurement on the detector is a projective measurement in the basis $\{|u_l^{m,\tau}\rangle_d\}_{l=0}^{n-1}$, such that the  identification of a detector state eliminates $N-m$ of the $N$ possible quanton path states. The resulting probability distribution $\{|\lambda_\ell|^2\}$ is given by Eq.~(\ref{eq:lambdasolution}), which yields the maximum which-path knowledge allowed by the quanton state's coherence, thus saturating the duality relation.

We now consider the general case of uniform detector states $|u_\ell\rangle_d$ with arbitrary support $\mathcal{I}$. The resulting quanton-detector reads
\begin{align}    \label{eq:Psi_maxentangled}
    |\Psi\rangle &= \frac{1}{\sqrt{N}}\sum_{\ell=0}^{N-1}|\ell\rangle_q|u_{\ell}\rangle_d = 
\frac{1}{\sqrt{n}}\sum_{k\in\mathcal{I}}|\varphi_k\rangle_q|k\rangle_d, 
\end{align}
where
\begin{align}    \label{eq:quanton_Fourier}
    |\varphi_k\rangle_q &= \frac{1}{\sqrt{N}}\sum_{\ell=0}^{N-1}\omega^{k\ell}|\ell\rangle_q 
\end{align}
are the Fourier transforms of the path basis states. The second expression in Eq.~(\ref{eq:Psi_maxentangled}) shows that, in this general case, $|\Psi\rangle$ is maximally entangled with Schmidt rank $n$; the state $|\Psi^{m,\tau}\rangle$ [Eq.~(\ref{eq:Psi_saturation})] is recovered when $\mathcal{I}=\mathcal{I}_{m,\tau}$. Although the state is maximally entangled, for $\mathcal{I}\neq\mathcal{I}_{m,\tau}$ it does not exhibit the block-structured correlations observed in the case of equally spaced supports: a projective measurement in the basis $\{|k\rangle_d\}_{k\in\mathcal{I}}$ leaves the quanton in a state of complete uncertainty in the path basis, namely $|\varphi_k\rangle_q$. 

The impossibility of saturation in this case stems from the fact that the detector states form an overcomplete (and hence linearly dependent) set in the $n$-dimensional subspace of the detector space. This precludes any measurement strategy from achieving the saturating probability distribution $\{|\lambda_\ell|^2\}$ of Eq.~(\ref{eq:lambdasolution}). In particular, for $n=N/m$ and $\mathcal{I}\neq\mathcal{I}_{m,\tau}$, the identification of $|u_\ell\rangle_d$ eliminates fewer than $N-m$ of the $N$ possible quanton path states, thus yielding less which-path knowledge than in the saturating case. Let us illustrate these aspects with the following example, introduced in Sec.~\ref{subsec:ResultsME}.

\subsubsection{Six-path interferometers with two-dimensional detector states}

In a uniform six-path interferometer with uniform two-dimensional detector states, the coherence of the quanton state is $\mathcal{C}=0.613$. In this scenario, three inequivalent classes of detector-state sets have been identified [see Fig.~\hyperref[fig:MELD]{\ref{fig:MELD}(d)}], each corresponding to a distinct support structure: equally spaced, unequally spaced adjacent, and unequally spaced nonadjacent. We start our analysis with the equally spaced support $I_{3,0}=\{0,3\}$, corresponding to the saturating set defined in Eq.~(\ref{eq:uniform_mtau}), in the subspace spanned by $\{|0\rangle_d,|3\rangle_d\}$. From Eqs.~(\ref{eq:Psi_saturation}) and (\ref{eq:psi_lq}), the quanton-detector state takes the form
\begin{multline}    \label{eq:Psi_30}
    |\Psi^{3,0}\rangle = \frac{1}{\sqrt{2}}\left[\left(\frac{|0\rangle_q+|2\rangle_q+|4\rangle_q}{\sqrt{3}}\right)|u_0^{3,0}\rangle_d \right. \\ + \left. \left(\frac{|1\rangle_q+|3\rangle_q+|5\rangle_q}{\sqrt{3}}\right)|u_1^{3,0}\rangle_d\right],    
\end{multline}
where $|u_l^{3,0}\rangle_d\propto|0\rangle_d+(-1)^l|3\rangle_d$. Since ${}_d\langle u_0^{3,0}|u_1^{3,0}\rangle_d=0$, the detector states are perfectly distinguishable; thus, identifying one of them eliminates three out of the six possible quanton path states. Using Eq.~(\ref{eq:Lambda}), we obtain the probability distribution $\left(\frac{1}{3},0,\frac{1}{3},0,\frac{1}{3},0\right)$ for $\{|\lambda_\ell|^2\}_{\ell=0}^{5}$, in agreement with Eq.~(\ref{eq:lambdasolution}). This yields $\mathcal{K}(\mathbf{\Pi}_{\scriptscriptstyle\mathrm{ME}})=0.387$, and hence $\mathcal{C}+\mathcal{K}(\mathbf{\Pi}_{\scriptscriptstyle\mathrm{ME}})=1$. The same conclusions apply to $\mathcal{I}_{3,1}=\{1,4\}$ and $\mathcal{I}_{3,2}=\{2,5\}$. 

Now we consider the unequally spaced adjacent support $\mathcal{I}=\{0,1\}$, which corresponds to a nonsaturating set of detector states in the subspace spanned by $\{|0\rangle_d,|1\rangle_d\}$. In this case, the quanton-detector state reads 
\begin{multline}   
    |\Psi\rangle = \frac{1}{\sqrt{6}} \Big(|0\rangle_q|u_{0_+}\rangle_d + |1\rangle_q|u_{1_+}\rangle_d + |2\rangle_q|u_{2_+}\rangle_d  \\ + |3\rangle_q|u_{0_-}\rangle_d + |4\rangle_q|u_{1_-}\rangle_d + |5\rangle_q|u_{2_-}\rangle_d  \Big), 
\end{multline}
where $|u_{l_\pm}\rangle_d\propto|0\rangle_d\pm\omega^l|1\rangle_d$ (for $l=0,1,2$), so that ${}_d\langle u_{l_+}|u_{l_-}\rangle_d=0$ while ${}_d\langle u_{l_\pm}|u_{l'_\pm}\rangle_d\neq 0$ for $l\neq l'$. Therefore, when the detector state is identified via ME measurement, e.g., as $|u_{l_+}\rangle_d$, the quanton path state cannot be $|l+3\rangle_q$, eliminating only one of the six possible alternatives. This reduces the which-path knowledge in comparison to the previous case. From Eq.~(\ref{eq:Lambda}), we obtain the probability distribution $\left(\frac{1}{3},\frac{1}{4},\frac{1}{12},0,\frac{1}{12},\frac{1}{4}\right)$ for $\{|\lambda_\ell|^2\}_{\ell=0}^{5}$, which yields $\mathcal{K}(\mathbf{\Pi}_{\scriptscriptstyle\mathrm{ME}})=0.178$ and $\mathcal{C}+\mathcal{K}(\mathbf{\Pi}_{\scriptscriptstyle\mathrm{ME}})=0.791$. The same conclusions hold for the other supports with this structure: $\mathcal{I}=\{1,2\},\{2,3\},\{3,4\},\{4,5\},$ and $\{0,5\}$. 

Finally, we examine the unequally spaced nonadjacent support $\mathcal{I}=\{0,2\}$, associated with a nonsaturating set of detector states in the subspace spanned by $\{|0\rangle_d,|2\rangle_d\}$. Here, the quanton-detector state is given by 
\begin{multline}      \label{eq:Psi_trine}
    |\Psi'\rangle = \frac{1}{\sqrt{3}}\left[\left(\frac{|0\rangle_q+|3\rangle_q}{\sqrt{2}}\right)|u_0\rangle_d + \left(\frac{|1\rangle_q+|4\rangle_q}{\sqrt{2}}\right)|u_1\rangle_d \right. \\ + \left. \left(\frac{|2\rangle_q+|5\rangle_q}{\sqrt{2}}\right)|u_2\rangle_d\right],    
\end{multline}
where $|u_{l}\rangle_d\propto|0\rangle_d+\omega^{2l}|2\rangle_d$ (for $l=0,1,2$), so that ${}_d\langle u_{l}|u_{l'}\rangle_d\neq 0$  for $l\ne l'$. Thus, the identification of the detector state does not eliminate any of the quanton path states, reducing the which-path knowledge in comparison to both previous cases. In this case, the probability distribution $\{|\lambda_\ell|^2\}_{\ell=0}^{5}$ is given by $\left(\frac{1}{3},\frac{1}{12},\frac{1}{12},\frac{1}{3},\frac{1}{12},\frac{1}{12}\right)$, which yields $\mathcal{K}(\mathbf{\Pi}_{\scriptscriptstyle\mathrm{ME}})=0.129$ and $\mathcal{C}+\mathcal{K}(\mathbf{\Pi}_{\scriptscriptstyle\mathrm{ME}})=0.742$. The same conclusions hold for the other supports with this structure: $\mathcal{I}=\{0,4\},\{1,3\},\{1,5\},\{2,4\},$ and $\{3,5\}$. This example will be revisited in what follows to analyze its implications for the saturation of the duality relation.

\subsection{Lack of universal saturation in multipath duality relations for pure states}

For two-path interferometers with pure quanton-detector states, Englert's duality relation in Eq.~(\ref{eq:DualityEnglert}) always saturates \cite{Englert1996}. In this context, when generalizing to multipath configurations, there was a natural expectation that the corresponding relations should also exhibit universal saturation for pure states \cite{Bimonte2003Comment,Bimonte2003}. However, this feature is not observed in either nonentropic \cite{Durr2001} or entropic \cite{Bagan20} generalizations, including the present work.

In the entropic framework considered here, it is worth noting that the two-path duality relation for pure states---where $\mathcal{K}_\textrm{max}=\mathcal{K}(\mathbf{\Pi}_{\scriptscriptstyle\mathrm{ME}})$---exhibits only trivial saturation (see Fig.~\ref{fig:CxK_N2}). This might suggest that the lack of universal saturation observed in the multipath case may originate from the entropic nature of the quantifiers involved or from the suboptimality of the employed measurement strategies, given that $\mathcal{K}_\textrm{max}\geqslant\mathcal{K}(\mathbf{\Pi}_{\scriptscriptstyle\mathrm{ME}})$. However, as we show below, this behavior is in fact rooted in the richer structure of the quanton-detector correlations and therefore goes beyond the specific framework of the duality relation.

To make this point explicit, we consider again the example discussed in the previous subsection, with the quanton-detector state $|\Psi'\rangle$ given by Eq.~(\ref{eq:Psi_trine}). In this case, the detector states lie on the equator of the Bloch sphere, with pairwise angular separation of $2\pi/3$. Thus, they correspond to the well-known trine states for which the POVM achieving accessible information was established in \cite{Sasaki99}. Using this result, we obtain $\mathcal{K}_\textrm{max}=0.226=1.75\times\mathcal{K}(\mathbf{\Pi}_{\scriptscriptstyle\mathrm{ME}})$, and hence $\mathcal{C}+\mathcal{K}_\textrm{max}=0.839$. Despite extracting the total available path information and improving over ME, this optimal measurement does not lead to saturation of the duality relation. 

A comparison between the maximally entangled states of Schmidt rank $n$, $|\Psi^{3,0}\rangle$ [Eq.~(\ref{eq:Psi_30})] and $|\Psi'\rangle$ [Eq.~(\ref{eq:Psi_trine})], shows that two quanton-detector states can yield the same quanton coherence $\mathcal{C}$ but different values of $\mathcal{K}_\textrm{max}$. In particular, $|\Psi^{3,0}\rangle$ leads to saturation of the duality relation, whereas $|\Psi'\rangle$ does not. This demonstrates that, in multipath interferometers, distinct correlation structures impose intrinsic limitations on attaining $\mathcal{K}_\textrm{max}=1-\mathcal{C}$. The lack of universal saturation thus follows directly from the richer structure of these correlations, rather than from the specific choice of quantifiers or measurement strategies.

\section{Conclusion}
\label{sec:Conclusion}

In this work, we developed a systematic quantitative investigation of wave-particle duality based on entropic trade-off relations between quantum coherence ($\mathcal{C}$) and which-path knowledge $[\mathcal{K}(\mathbf{\Pi})]$ in uniform multipath interferometers with symmetric detector states. 

For general interferometric frameworks, characterized by nonuniformity and arbitrary detector states, one is restricted to upper bounds for $\mathcal{K}(\mathbf{\Pi})$. As discussed earlier, this prevents a precise assessment of the tightness of the resulting relations, their dependence on the employed detector-state discrimination strategy, and the identification of genuine saturation. This upper-bound-based approach has been adopted in recent works with which-path knowledge acquired via UD \cite{Bera15}, ME \cite{Bagan16}, and standard strategies that interpolate between them \cite{Bagan20}.

Our framework, on the other hand, although less general, allowed us to derive duality relations with exact $\mathcal{K}(\mathbf{\Pi})$ across multiple detector-state discrimination regimes, including those studied in Refs.~\cite{Bera15,Bagan16,Bagan20} and extending beyond them through the concatenated measurement. We thus identified a strategy-dependent hierarchy that generally emerges in the tightness of the resulting relations. For a fixed $\xi\in(0,1]$, concatenated FRIO always yields a relation at least as tight as its standard counterpart, whereas ME ($\xi=0$) typically yields the tighter relation. Although the latter ordering can be violated for particular detector-state ensembles, these exceptions do not establish a general advantage over ME when the full $\mathcal{C}$--$\mathcal{K}(\mathbf{\Pi})$ regions are considered. Overall, the relations generally become less tight as $\xi$ increases, with the standard FRIO relation at the limiting case $\xi=1$ (UD or MC measurements) typically yielding the least tight one. Thus, for linearly independent detector states, while the UD upper bound for the which-path knowledge yields a formally saturated relation \cite{Bera15}, our exact quantification shows that such saturation does not, in general, correspond to an attainable limit of the trade-off.

We further identified the saturation of the duality relations within our framework, demonstrating that it occurs when the detector states form a Fourier basis for an $n$-dimensional subspace of the detector space, with $n|N$. In this case, we have shown that the quanton-detector state is maximally entangled with Schmidt rank $n$ and exhibits a block structure in the path basis. By exploiting this result, we explain the lack of universal saturation in multipath duality relations for pure states as a direct consequence of the underlying correlation structure.

From an operational standpoint, our results also highlight the role of the number of paths in the design of uniform multipath interferometric setups with symmetric detector states. While the duality relations hold universally, the nonprime-$N$ condition emerges as the precise algebraic requirement for observing nontrivial saturation. 

The discussion of wave-particle duality dates back almost a century, yet it remains relevant, particularly in the context of duality relations in multipath interferometers. Our findings offer novel insights into this topic and pave the way for experimental verification. Such a test would require the preparation of the correlated quanton-detector state considered in this work together with the implementation of the corresponding detector-state discrimination strategy. Since discrimination protocols for symmetric states have already been demonstrated experimentally using photonic qudits encoded in transverse path modes \cite{Prosser17, Prosser22, Melo23}, our results should be amenable, in principle, to experimental investigation in multipath photonic platforms capable of engineering high-dimensional quantum states. Examples of suitable platforms include integrated photonic circuits and optical architectures based on transverse modes of light, such as path and orbital-angular-momentum encodings.

\begin{acknowledgments}
This work was supported by CNPq INCT-IQ Grant No.\ 465469/2014-0 and CNPq Grants No.\ 422300/2021-7 and No.\ 303212/2022-5. L. F. M. acknowledges financial support from CAPES -- Finance Code 001, and from the projects ``Receptores n\~ao-convencionais em CV-QKD'', ``An\'alise de seguran\c{c}a de protocolos P\&M CV-QKD'' and ``Design e arquiteturas de protocolos CV-QKD'' supported by the EMBRAPII CIMATEC Competence Center in Quantum Technologies -- Quantum Industrial Innovation (QuIIN), with financial resources from the PPI IoT/Manufatura 4.0 of the MCTI grant number 053/2023, signed with EMBRAPII. O. J. was supported by an internal grant from Universidad Mayor (PUENTE-2024-17).
\end{acknowledgments}

\appendix

\section{Numerical analysis of the monotonicity of \texorpdfstring{$\mathcal{K}(\mathbf{\Pi}_{\scriptscriptstyle\mathrm{FRIO}})$}{K(Pi{FRIO})} with \texorpdfstring{$\xi$}{xi}}    \label{app:MonotonicityFRIO}

In this Appendix, we investigate the behavior of the which-path knowledge for the standard FRIO measurement [$\mathcal{K}(\mathbf{\Pi}_{{\scriptscriptstyle\mathrm{FRIO}}})$ in Eq.~(\ref{eq:Kfrio})] as a function of the separation parameter $\xi$. For $N=2$, it can be proved analytically that $d\mathcal{K}/d\xi\leqslant 0$ for $a_\textrm{min}^2\in[0,1/2]$ and $\xi\in[0,1]$, with equality holding only for the trivial cases of identical or orthogonal detector states. For $N>2$, however, an analytical treatment for a fixed set of detector states specified by arbitrary coefficients $\{a_k\}$ becomes substantially more involved. We therefore performed a systematic numerical investigation to assess the behavior of $\mathcal{K}(\mathbf{\Pi}_{{\scriptscriptstyle\mathrm{FRIO}}})$, restricting our analysis, without loss of generality, to linearly independent states ($n=N$).

Our numerical exploration using randomly generated sets $\{a_k\}$ in low dimensions ($N=3,\ldots,6$) detected violations of $d\mathcal{K}/d\xi\leqslant 0$ only for a restricted family of symmetric states, characterized by one dominant coefficient and an approximately degenerate set of $N-1$ smaller coefficients. A more in-depth investigation into this family revealed that the observed violations are progressively suppressed as the smaller coefficients become increasingly nondegenerate, whereas the largest violations were found for the degenerate case, i.e.,
\begin{equation}
\label{eq:2amp_family}
\{a_k^2\}
=
\left\{q,\frac{1-q}{N-1},\ldots,\frac{1-q}{N-1}\right\}
\end{equation}
with $q\in[1/N,1]$. In the following, we present our numerical analysis focusing on this family of symmetric states, which we refer to as the $1$-max two-amplitude family.\footnote{More generally, a two-amplitude family of symmetric states has $d$ degenerate maximum amplitudes (and hence $N-d$ degenerate minimum amplitudes), for $1\leqslant d\leqslant N-1$, as defined in Ref.~\cite{Herzog12} in the context of state discrimination. We refer to it here as a $d$-max two-amplitude family.} 

First, we define two gain functions that depend on the ensemble parameter $q$:
\begin{align}
\mathcal{G}_{\scriptscriptstyle\mathrm{FRIO}}(q) &= \max_\xi\, [\mathcal{K}(\mathbf{\Pi}_{{\scriptscriptstyle\mathrm{FRIO}}})] - \mathcal{K}(\mathbf{\Pi}_{{\scriptscriptstyle\mathrm{ME}}}), \\
\mathcal{G}_{\scriptscriptstyle\mathrm{UD}}(q) &= \mathcal{K}(\mathbf{\Pi}_{{\scriptscriptstyle\mathrm{UD}}}) - \mathcal{K}(\mathbf{\Pi}_{{\scriptscriptstyle\mathrm{ME}}}).
\end{align}
Here, $\mathcal{G}_{\scriptscriptstyle\mathrm{FRIO}}(q)>0$ indicates that, for a given $q$, at least one FRIO measurement yields more which-path knowledge than ME. On the other hand, $\mathcal{G}_{\scriptscriptstyle\mathrm{UD}}(q)>0$ indicates an advantage of UD over ME. 

%%%%%%%%%%%%%% FIGURE 6 %%%%%%%%%%%%%%%%%%%%
\begin{figure*}[!t]
    \centering
    \includegraphics[width=.979\textwidth]{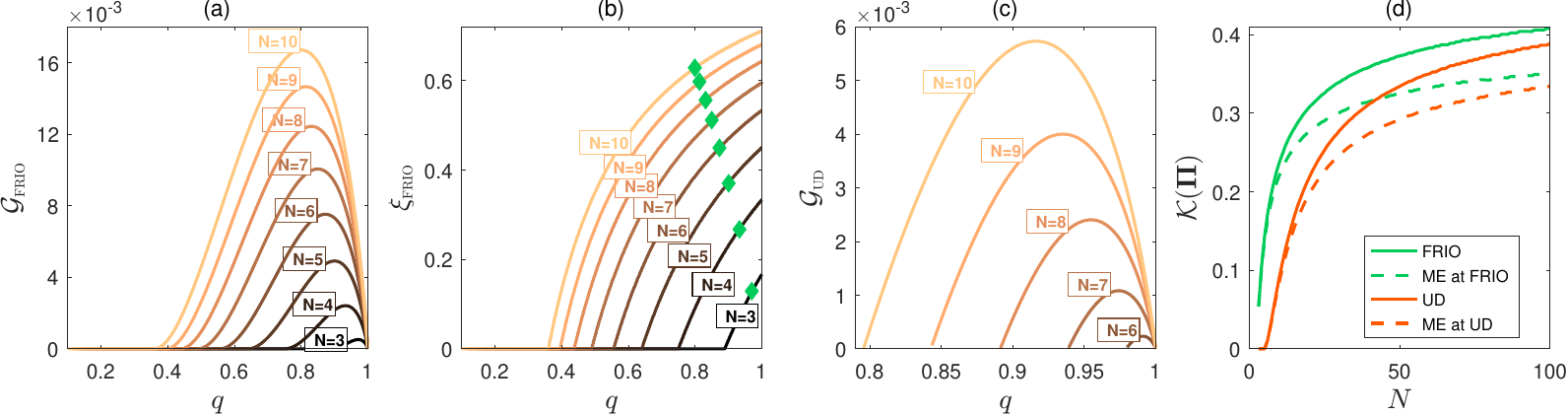}
\caption{Numerical characterization of the behavior of $\mathcal{K}(\mathbf{\Pi}_{{\scriptscriptstyle\mathrm{FRIO}}})$ for the $1$-max two-amplitude family of symmetric detector states [Eq.~(\ref{eq:2amp_family})]. (a) $\mathcal{G}_{{\scriptscriptstyle\mathrm{FRIO}}}$ vs $q$ for $N$ shown in the labels. (b) $\xi_{{\scriptscriptstyle\mathrm{FRIO}}}$ vs $q$; the green diamonds indicate the values of $\xi_{{\scriptscriptstyle\mathrm{FRIO}}}$ corresponding to the ensembles that maximize $\mathcal{G}_{{\scriptscriptstyle\mathrm{FRIO}}}$. (c) Positive values of $\mathcal{G}_{{\scriptscriptstyle\mathrm{UD}}}(q)$. (d) Which-path knowledge vs $N$, evaluated at the ensembles that maximize $\mathcal{G}_{{\scriptscriptstyle\mathrm{FRIO}}}$ (green) and $\mathcal{G}_{{\scriptscriptstyle\mathrm{UD}}}$ (orange). Solid and dashed lines denote FRIO/UD and ME, respectively.}
\label{fig:MonotonicityTest}
\end{figure*}
%%%%%%%%%%%%%%%%%%%%%%%%%%%%%%%%%%%%%%%%

For dimensions in the range of $N=3$ to $N=100$, we numerically evaluate not only the gain functions defined above, but also additional quantities that characterize the behavior of $\mathcal{K}(\mathbf{\Pi}_{{\scriptscriptstyle\mathrm{FRIO}}})$ for the $1$-max two-amplitude family of detector states. Figure~\hyperref[fig:MonotonicityTest]{\ref{fig:MonotonicityTest}(a)} shows $\mathcal{G}_{\scriptscriptstyle\mathrm{FRIO}}(q)$ for $N=3,\ldots,10$. It can be observed that both the range of $q$ values yielding a positive gain and the magnitude of the gains increase with $N$; the occurrence of $\mathcal{G}_{\scriptscriptstyle\mathrm{FRIO}}>0$ demonstrates the violation of the monotonicity condition $d\mathcal{K}/d\xi\leqslant 0$ for $N>2$. In Fig.~\hyperref[fig:MonotonicityTest]{\ref{fig:MonotonicityTest}(b)}, we plot
$\xi_{\scriptscriptstyle\mathrm{FRIO}}(q) = \arg\max_\xi\,[\mathcal{K}(\mathbf{\Pi}_{{\scriptscriptstyle\mathrm{FRIO}}})]$, which determines the FRIO strategy associated with $\mathcal{G}_{\scriptscriptstyle\mathrm{FRIO}}(q)$. Clearly, the range of $\xi_{\scriptscriptstyle\mathrm{FRIO}}$ associated with positive gains also broadens with $N$. In addition, the green diamond markers in Fig.~\hyperref[fig:MonotonicityTest]{\ref{fig:MonotonicityTest}(b)} indicate the corresponding $\xi_{\scriptscriptstyle\mathrm{FRIO}}$ for the ensemble that maximizes $\mathcal{G}_{\scriptscriptstyle\mathrm{FRIO}}$; these markers shift toward the UD limit $\xi=1$ as $N$ increases. In Fig.~\hyperref[fig:MonotonicityTest]{\ref{fig:MonotonicityTest}(c)}, we plot the values of $\mathcal{G}_{\scriptscriptstyle\mathrm{UD}}>0$. Our tests showed that these occurred only for $N>4$.\footnote{For $N=5$, the maximum $\mathcal{G}_{\scriptscriptstyle\mathrm{UD}}\sim 10^{-9}$ and is not shown in the plot.} Similarly to $\mathcal{G}_{\scriptscriptstyle\mathrm{FRIO}}$, both the range of $q$ values yielding a positive gain and the magnitude of the gains grow with $N$, although more slowly.

Figure~\hyperref[fig:MonotonicityTest]{\ref{fig:MonotonicityTest}(d)} shows the values of the which-path knowledge as a function of $N$, for $N=3,\ldots,100$, obtained as follows. For each dimension, the values of $q$ that maximize the gains $\mathcal{G}_{\scriptscriptstyle\mathrm{FRIO}}$ and $\mathcal{G}_{\scriptscriptstyle\mathrm{UD}}$ are first identified independently. The corresponding values of $\mathcal{K}(\mathbf{\Pi}_{{\scriptscriptstyle\mathrm{FRIO}}})$ and $\mathcal{K}(\mathbf{\Pi}_{{\scriptscriptstyle\mathrm{ME}}})$ are then evaluated at the former, while $\mathcal{K}(\mathbf{\Pi}_{{\scriptscriptstyle\mathrm{UD}}})$ and $\mathcal{K}(\mathbf{\Pi}_{{\scriptscriptstyle\mathrm{ME}}})$ are evaluated at the latter. The results show that these $\mathcal{K}$s increase with the dimension, while the values obtained with FRIO and UD progressively approach each other and become increasingly separated from the corresponding ME values. This behavior is consistent with the increase of the optimal separation parameter toward the UD limit observed in Fig.~\hyperref[fig:MonotonicityTest]{\ref{fig:MonotonicityTest}(b)}.

Overall, these results provide a counterexample to the expectation that $\mathcal{K}(\mathbf{\Pi}_{{\scriptscriptstyle\mathrm{FRIO}}})$ is nonincreasing with $\xi$ for $N>2$ (as it is for $N=2$). However, numerical evidence indicates that this behavior is not generic within the broader class of symmetric detector states addressed here, but is restricted to a specific subclass encompassing the $1$-max two-amplitude family and small deviations from it. Thus, the advantage of FRIO or UD in this scenario should be understood as a property of this particular family rather than as an indication that these strategies provide more which-path knowledge than the ME measurement for general symmetric states. To illustrate this point, we plot $\mathcal{C}$ versus $\mathcal{K}(\mathbf{\Pi}_{{\scriptscriptstyle\mathrm{FRIO}}})$ in Fig.~\ref{fig:CKfrio} for the $1$-max two-amplitude family (cyan solid lines in the first and third rows). Clearly, these curves correspond to the upper boundaries of the regions spanned by the $(\mathcal{K},\mathcal{C})$ points in each case, meaning that this family yields the tightest relations among the symmetric states for standard FRIO or UD. More importantly, the curves associated with FRIO (first row) and UD (third row) lie, to a large extent, within the region populated by the general symmetric states when ME is used (boundaries defined by the black solid lines). Small excursions outside the ME region are observed only at high coherence values, corresponding to large values of $q$, i.e., nearly parallel detector states.

To conclude this appendix, we note that the detector states in the $1$-max two-amplitude family [Eq.~(\ref{eq:2amp_family})] have equal mutual overlaps, namely,  $\langle\alpha_\ell|\alpha_{\ell'}\rangle=\frac{Nq-1}{N-1}$, for $\ell\neq\ell'$. These are precisely the states considered in Ref.~\cite{Bagan20} for which the entropic duality relation saturates in the asymptotic limit $N\rightarrow\infty$ when which-path information is obtained using UD.\footnote{Among the $d$-max two-amplitude families, the only other case yielding symmetric states with constant mutual overlaps corresponds to $d=N-1$, with fiducial coefficients $\{a_\textrm{max},\ldots,a_\textrm{max},a_\textrm{min}\}$. In this case, however, the limit $a_\textrm{min}\to 0$ leads to detector states that encode path information, rather than to the parallel states obtained for the $1$-max family in Eq.~(\ref{eq:2amp_family}). No violation of the monotonicity condition was observed for this family in our numerical tests.} Our numerical results are consistent with this asymptotic behavior, as the which-path knowledge obtained with UD and FRIO progressively approaches the same value as $N$ increases.

%%%%%%%%%%%%%%%%%%% APPENDIX B %%%%%%%%%%%%%%%%%%%%%%%%%

\section{Comparison between \texorpdfstring{$\mathcal{K}(\mathbf{\Pi}_{\mathrm{conc}})$}{K(Pi{conc})}  and \texorpdfstring{$\mathcal{K}(\mathbf{\Pi}_{\scriptscriptstyle\mathrm{ME}})$}{K(Pi{ME})} }
\label{app:conc_me}

In the concatenated FRIO strategy, the which-path knowledge is given by Eq.~(\ref{eq:Kconc}), where the second term is the information extracted from the failure branch. Since this contribution is nonnegative, concatenation always satisfies $\mathcal{K}(\mathbf{\Pi}_{\mathrm{conc}})\geqslant \mathcal{K}(\mathbf{\Pi}{\scriptscriptstyle_\mathrm{FRIO}})$. On the other hand, the comparison of the which-path knowledge obtained from concatenated FRIO and ME is more subtle. Despite the natural ordering provided by the minimum-error rate criterion, namely, $P_e^{\scriptscriptstyle\mathrm{ME}}\leqslant P_e^{\mathrm{conc}}$, this relation does not immediately imply $\mathcal{K}(\mathbf{\Pi}_{\scriptscriptstyle\mathrm{ME}})\geqslant \mathcal{K}(\mathbf{\Pi}_{\mathrm{conc}})$. To examine whether this latter ordering can be violated, we define
\begin{align}    \label{eq:delta_conc}
\Delta_{\mathrm{conc}}(\xi) &=
\mathcal{K}(\mathbf{\Pi}_{\mathrm{conc}}) -
\mathcal{K}(\mathbf{\Pi}_{\scriptscriptstyle\mathrm{ME}}) \nonumber\\
&= \left[\mathcal{K}(\mathbf{\Pi}_{\scriptscriptstyle\mathrm{FRIO}}) -
\mathcal{K}(\mathbf{\Pi}_{\scriptscriptstyle\mathrm{ME}})\right] + P_f(\xi)\widetilde{\mathcal{H}}(\{p^f_{\ell|0}\}) \nonumber\\
&\equiv \Delta_{\scriptscriptstyle\mathrm{FRIO}}(\xi) + P_f(\xi)\widetilde{\mathcal{H}}(\{p^f_{\ell|0}\}).
\end{align}
In the last expression, the first term measures the gain or loss of the standard FRIO strategy relative to ME, whereas the second accounts for the additional information recovered from the failure states.

This decomposition allows two different mechanisms leading to $\Delta_{\mathrm{conc}}>0$ to be distinguished. First, whenever $\Delta_{\scriptscriptstyle\mathrm{FRIO}}>0$, the ordering is already violated by the standard FRIO strategy, and the nonnegative contribution from the failure branch can only increase this difference. We refer to such cases as inherited violations, which therefore include those arising from the detector states discussed in the previous appendix. The second mechanism corresponds to the case where $\Delta_{\scriptscriptstyle\mathrm{FRIO}}\leqslant 0$ and 
\begin{align}       \label{eq:proper_violation}
P_f(\xi)\widetilde{H}(\{p^f_{\ell|0}\})  &> \mathcal{K}(\mathbf{\Pi}_{\scriptscriptstyle\mathrm{ME}}) - \mathcal{K}(\mathbf{\Pi}_{\scriptscriptstyle\mathrm{FRIO}})= -\Delta_{\scriptscriptstyle\mathrm{FRIO}}.
\end{align}
We refer to these cases as proper violations associated with the concatenated measurement. They occur when the information recovered from the failure branch is sufficiently large to compensate for the loss of the standard FRIO strategy relative to ME.

Numerical searches over randomly generated sets $\{a_k\}$ in dimensions up to $N=10$ show that proper violations do occur. However, unlike the FRIO case, these searches reveal a considerably more diverse structure of the corresponding ensembles. Violations were found, for instance, for states characterized by (i) one dominant coefficient and an approximately degenerate set of $N-1$ smaller coefficients, as in the FRIO case;\footnote{In contrast to standard FRIO, the exactly degenerate $1$-max two-amplitude family does not produce proper violations, since the corresponding failure states are parallel and therefore carry no which-path information.} (ii) reflection-symmetric distributions modulo $N$ with two dominant coefficients, two intermediate coefficients, and a degenerate set of $N-4$ smaller coefficients, found for $N=5$ and $7$; and (iii) the 2-max two-amplitude family for $N=6,8$, and $10$, and the 3-max two-amplitude family for $N=9$, with equally spaced maximum coefficients. In general, the largest proper violations found in these searches were of the order of $10^{-3}$.

Although Eq.~(\ref{eq:proper_violation}) provides an exact condition for a proper violation, the results above do not point to a simple structural characterization of the ensembles satisfying it. In particular, the occurrence of violations was not found to be determined solely by properties such as the number of degenerate amplitudes or their arrangement within the fiducial state. This difficulty can be understood from Eq.~(\ref{eq:proper_violation}), since a proper violation depends on the balance between $P_f(\xi)\widetilde{\mathcal{H}}(\{p^f_{\ell|0}\})$ and $\mathcal{K}(\mathbf{\Pi}_{\scriptscriptstyle\mathrm{ME}}) -\mathcal{K}(\mathbf{\Pi}_{\scriptscriptstyle\mathrm{FRIO}})$. Consequently, neither a large $\widetilde{\mathcal{H}}(\{p^f_{\ell|0}\})$ nor a large $P_f(\xi)$ is by itself sufficient to determine whether it occurs. A complete characterization of the ensembles producing proper violations therefore appears to require a more detailed analysis beyond the scope of the present work.

It is important, however, to distinguish this ensemble-by-ensemble comparison from the geometry of the full $\mathcal{C}$--$\mathcal{K}(\mathbf{\Pi})$ region. Rather than attempting to identify all violating ensembles, a more global question for the present analysis is whether these violations extend the upper boundary of the region attainable with ME measurements for the general symmetric states considered here. The condition $\mathcal{K}(\mathbf{\Pi}_{\mathrm{conc}})>\mathcal{K}(\mathbf{\Pi}_{\scriptscriptstyle\mathrm{ME}})$ compares the two strategies for the same detector-state ensemble, but does not imply that the corresponding point $(\mathcal{K}(\mathbf{\Pi}_{\mathrm{conc}}),\mathcal{C})$ lies above the ME boundary. Indeed, the scatter plots in the second and fourth rows of Fig.~\ref{fig:CKfrio} provide no visual evidence of such an extension.

%%%%%%%%%%%%%% FIGURE 7 %%%%%%%%%%%%%%%%%%%%
\begin{figure}[!t]
    \centering
    \includegraphics[width=.8\columnwidth]{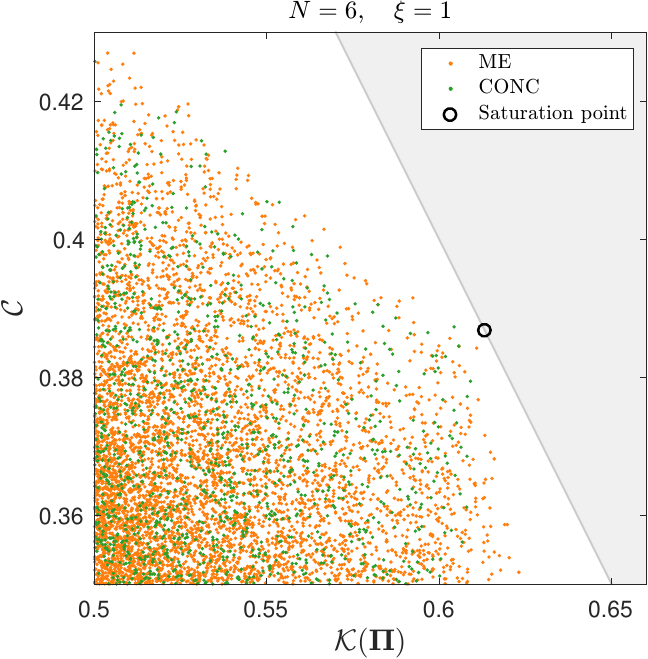}
\caption{Enlarged view of the $\mathcal{C}$--$\mathcal{K}(\mathbf{\Pi})$ region for $N=6$ and $\xi=1$, showing the data for ME (orange dots) and concatenated FRIO (green dots) in the vicinity of the nontrivial saturation point (open circle). The sampling becomes increasingly sparse toward the upper edge of the accessible region, particularly in the neighborhood of the saturation point.}
\label{fig:ZoomME_conc}
\end{figure}
%%%%%%%%%%%%%%%%%%%%%%%%%%%%%%%%%%%%%%%%

To refine this observation, we performed a more quantitative numerical analysis of the upper boundary of the ME region. For each point $(\mathcal{K}(\mathbf{\Pi}_{\mathrm{conc}}),\mathcal{C})$, we constructed a local estimate of the ME upper boundary using the $M$ sampled ME points closest in coherence and compared $\mathcal{K}(\mathbf{\Pi}_{\mathrm{conc}})$ with the largest value of $\mathcal{K}(\mathbf{\Pi}_{\scriptscriptstyle\mathrm{ME}})$ within this local set.\footnote{In sparsely sampled regions, the largest value of $\mathcal{K}(\mathbf{\Pi}_{\scriptscriptstyle\mathrm{ME}})$ within a finite random sample may underestimate the actual ME boundary. Consequently, small apparent excursions of concatenated FRIO points above the sampled ME envelope cannot by themselves be identified as genuine boundary crossings.} The analysis was repeated for increasing values of $M$ to assess the sensitivity of the comparison to the local sampling density. Increasing $M$ increases the statistical sampling while also enlarging the coherence interval over which the boundary is estimated.

For $N=3$, no points above the estimated ME boundary were found over the range of $M$ considered. For $N=4$ and $5$, small apparent excursions observed for smaller local samples disappeared as $M$ was increased. The case $N=6$ proved more delicate: apparent excursions persisted over a wider range of $M$, although both their number and magnitude decreased as the local sample was enlarged. Inspection of the ME data shows that these residual excursions are associated with regions where the sampling becomes increasingly sparse near the upper boundary. This effect is particularly pronounced around the nontrivial saturation points. An enlarged view of the region near the saturation point $\mathcal{C}\simeq 0.387$ is shown as an example in Fig.~\ref{fig:ZoomME_conc}. Therefore, the observed deviations from the estimated ME boundary cannot be reliably identified as genuine boundary crossings.

In summary, the relation
$\mathcal{K}(\mathbf{\Pi}_{\mathrm{conc}}) \leqslant \mathcal{K}(\mathbf{\Pi}_{\scriptscriptstyle\mathrm{ME}})$
does not hold universally on an ensemble-by-ensemble basis. Violations may either be inherited from standard FRIO or arise specifically from the information recovered from the failure branch. In the latter case, their occurrence is governed by the competition expressed in Eq.~(\ref{eq:proper_violation}). Nevertheless, despite these individual violations, our numerical results indicate that concatenated FRIO measurements do not extend the upper boundary of the $\mathcal{C}$--$\mathcal{K}$ region attainable with ME measurements within the general symmetric detector-state ensembles considered in this work.

%\bibliographystyle{apsrev4-2}
%\bibliography{refs}
\input{RefsWP.bbl}

\end{document}